%% file: zc3900.tex
\documentclass[3p,times,sort&compress]{elsarticle}
\include{quattro-sym}

\graphicspath{{figures/}}
\begin{document}
 
%%%%%%%%%%%%%%%%%%%%%%%%%%%%%%%%
%	Title
%%%%%%%%%%%%%%%%%%%%%%%%%%%%%%%%%
\title{Amplitude analysis and the nature of the $Z_c(3900)$}
%
%%%%%%%%%%%%%%%%%%%%%%%%%%%%%%%%%
%%	Author list
%%%%%%%%%%%%%%%%%%%%%%%%%%%%%%%%%
%
%
\author[jlab]{A.~Pilloni}
\ead{pillaus@jlab.org}
\author[UNAM]{C.~Fern\'andez-Ram\'irez}
\author[CEEM,IU]{A.~Jackura}
\author[CEEM,IU]{V.~Mathieu}
\author[bonn]{M.~Mikhasenko}
\author[ghent]{J.~Nys}
\author[jlab,CEEM,IU]{A.~P.~Szczepaniak}
\author{\\ \vspace{.4cm}(JPAC Collaboration)}

\address[jlab]{Theory Center, Thomas Jefferson National Accelerator Facility,
12000 Jefferson Avenue, Newport News, VA 23606, USA}
\address[UNAM]{Instituto de Ciencias Nucleares, Universidad Nacional Aut\'onoma de M\'exico, Ciudad de M\'exico 04510, Mexico}
\address[CEEM]{Center for Exploration of Energy and Matter, Indiana University, Bloomington, IN 47403, USA}
\address[IU]{Physics Department, Indiana University, Bloomington, IN 47405, USA}
\address[bonn]{Universit\"at Bonn, Helmholtz-Institut f\"ur Strahlen- und Kernphysik, 53115 Bonn, Germany}
\address[ghent]{Department of Physics and Astronomy, Ghent University, B-9000 Ghent, Belgium}

%%%%%%%%%%%%%%%%%%%%%%%%%%%%%%%%%
%%	Preprint Numbers
%%%%%%%%%%%%%%%%%%%%%%%%%%%%%%%%%
%%\preprint{JLAB-THY-16-}
%
%%%%%%%%%%%%%%%%%%%%%%%%%%%%%%%%%
%%	Collaboration
%%%%%%%%%%%%%%%%%%%%%%%%%%%%%%%%%
%%\collaboration{Joint Physics Analysis Center}
%
%%%%%%%%%%%%%%%%%%%%%%%%%%%%%%%%%
%%	Abstract  & PACS
%%%%%%%%%%%%%%%%%%%%%%%%%%%%%%%%%
\begin{abstract}
The microscopic nature of the \XYZ states remains an unsettled topic. 
We show how a thorough amplitude analysis of the data can help constraining  models of these states. Specifically, we consider the case of the $Z_c(3900)$ peak  and discuss possible scenarios of a QCD state, virtual state, or a  kinematical enhancement. We conclude that current 
data are not precise enough to distinguish between these hypotheses, however, the method we propose, when applied to the forthcoming high-statistics measurements should shed light on the nature of these exotic enhancements. 
\end{abstract}
\begin{keyword}
\PACS 14.40.Rt \sep 14.40.Pq \sep 11.55.Fv \sep 11.80.Et\\
JLAB-THY-16-2410
\end{keyword}
%%\date{\today}
\maketitle
%%%%%%%%%%%%%%%%%%%%%%%%%%%%%%%%
%	Introduction
%%%%%%%%%%%%%%%%%%%%%%%%%%%%%%%%

 The nature of the recently discovered \XYZ states remains a mystery, as they are at odds with the standard 
    quarkonium phenomenology. Most of the literature interprets these structures as multi-quark states~\cite{Maiani:2004vq,Faccini:2013lda,Maiani:2014aja,Brodsky:2014xia,Esposito:2015fsa}, loosely bound hadron molecules~\cite{Tornqvist:1993ng,Braaten:2003he,Close:2003sg,Swanson:2006st}, hybridized states~\cite{Esposito:2016itg,physrept}, hadroquarkonia~\cite{Dubynskiy:2008mq,Li:2013ssa}, or gluonic excitations~\cite{Guo:2008yz,Kou:2005gt}, or rescattering effects~\cite{Swanson:2014tra,Szczepaniak:2015eza} (criticized in~\cite{Guo:2014iya}); for a review, see~\cite{Chen:2016qju,Esposito:2014rxa,Brambilla:2014jmp,Lebed:2016hpi,physrept}. It is worth noticing that most of the \XYZ phenomena occur in a mass region where  there is an abundance of open channels, which potentially can result in 
  virtual state poles or anomalous thresholds.  In this letter we examine whether  existing data on the charged charmonium-like $Z_c(3900)$ enhancement can discriminate or not between these scenarios. 
 
 The $Z_c(3900)$ was discovered simultaneously by \bes~\cite{Ablikim:2013mio} and \belle~\cite{Liu:2013dau}. \bes observed
 an enhancement in the $\jpsi\,\pi$ mass distributions~\footnote{The charge conjugated modes are always understood.} of the reaction $e^+ e^- \to  \jpsi\,\pi^+ \pi^-$.   The center of mass energy was fixed at $E_\text{CM} = 4260\mev$, which matches with 
 the mass of the $Y(4260)$, leading to the possibility for the reaction to be dominated by $e^+ e^- \to  Y(4260) \to 
    J/\psi\, \pi^+ \pi^- $. \belle performed the analysis of the same final state with additional Initial State Radiation (ISR), $e^+ e^- \to  \gamma_\text{ISR} Y(4260) \to \gamma_\text{ISR} \jpsi \,\pi^+\pi^-$. 
    \bes observed a similar structure in the $\bar D D^*$ mass projection, in 
     the $e^+ e^- \to \bar D D^* \pi$ reaction~\cite{Ablikim:2013xfr,Ablikim:2015swa}. Evidence of a neutral isospin partner has been found by \bes and by an analysis of CLEO-$c$ data~\cite{Xiao:2013iha,Ablikim:2015gda,Ablikim:2015tbp}. The state has not been found either in $B$ decays~\cite{Chilikin:2014bkk}, or in photoproduction off protons~\cite{Adolph:2014hba}. 
  
  In the original analyses, the peak in the  $3900\mev$ mass region was assumed to be a resonance and was fitted with a Breit-Wigner formula 
   modified by a smooth background. Several authors considered alternative descriptions, in particular   emphasizing the role of singularities other than resonance poles. For example, in~\cite{Wang:2013cya} the $\jpsi\,\pi\pi $ Dalitz distribution was analyzed in  a model containing both, an anomalous threshold and a resonance. The anomalous threshold, 
 which originates from cross-channel exchanges, 
     leads to  a second-sheet singularity of a partial wave and produces a cusp-like enhancement on the real axis. Without sufficient resolution, anomalous threshold cusps may resemble Breit-Wigner distributions. 
       The authors  of~\cite{Wang:2013cya} model the interaction between the  $J/\psi\,\pi $ and the $D \bar D^*$  by the exchange of a cross-channel  $D_1(2420)$, which is a good candidate to create an anomalous cusp.
       The prediction of the model was compared with  the $\pi \pi$ and $\jpsi\,\pi$ spectra of the  $J/\psi\,\pi^+ \pi^- $ decay mode. The authors conclude that the cusp alone is not sufficient to describe the $Z_c(3900)$ peak and argue in favor of a resonance,  although no quantitive measures are given.  
 Numerous other works on cusps and/or poles typically assume a particular scenario for producing peaks 
 and compare model predictions to a subset of available data~\cite{Chen:2013coa,Wang:2013hga,Swanson:2014tra,Guo:2014iya,Aceti:2014uea,Szczepaniak:2015eza,Albaladejo:2015lob}.
 
  Given that the available published  data are not corrected for acceptance or efficiencies, and there is no polarization information, it is difficult to make a case for  a systematic fit of all the datasets.  Nevertheless, we will attempt such an analysis.   On the theoretical side, we use several parametrizations of the amplitudes which focus on the role of various singularities, without entering into the details of which model would be able to describe  their microscopic origin. 
     
 \section{Amplitude model}
 \label{sec:amplitude}
\begin{figure}[t]
 \begin{subfigure}[t]{.48\columnwidth}
 \centering
 \includegraphics[width=.4\columnwidth]{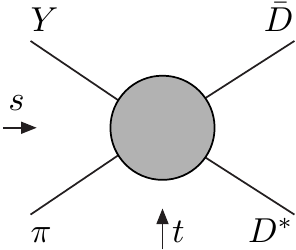}
 \caption{\small Channel $1$}
 \end{subfigure}
  \begin{subfigure}[t]{.48\columnwidth}
 \centering
 \includegraphics[width=.4\columnwidth]{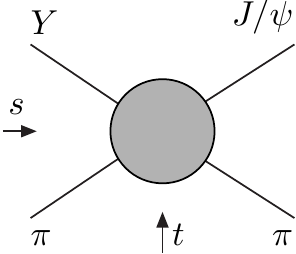}
 \caption{\small Channel $2$}
 \end{subfigure}
 \caption{\small Channel definitions. In channel $1$ we consider the exchange of a $D_1(2420)$ in $t$ and of a $\bar D_0(2400)$ in $u$ in addition to the possible $Z_c$ in $s$.  In channel $2$ we consider the exchange of a $f_0(980)$ and a $\sigma$ in $t$, in addition to the possible $Z_c$ in $s$ and $u$. }
 \label{fig:diagrams}
  \end{figure}

    Consider the three-body decay 
  $A\to BCD$. Under special kinematic conditions~\cite{Coleman:1965xm}, a cusp in the mass distribution of $BC$ can be generated, if there is 
    another available direct channel and if a resonance occurs in one of the two crossed channels near the physical 
     region~\cite{Anisovich:1995ab,Szczepaniak:2015hya}. In the absence of a coupled channel, the 
       crossed channel resonances lead to an enhancement in the Dalitz plot, which cancels out upon mass projection~\cite{schmid}. Such cusps are part of the production amplitude, {\it aka} left hand side branch points of partial waves.   In addition to this, partial waves have direct channel (right hand)  singularities, like threshold branch points,  or 
        virtual or resonance poles. 
        The definition of the channels relevant to this analysis is given in \figurename{~\ref{fig:diagrams}}.
The peak at $\sqrt{s} \sim 3900\mev$ may thus originate from a true $s$-channel resonance pole (the $Z_c$), a virtual state, the left hand branch point,  or a combination of both. 
 The best candidate to produce a triangle cusp is the $D_1(2420)$  resonance in the $t$-channel process $Y\, D \to \pi  D^*$. We also consider other possible exchanges, like the $D_0(2400)$ in $Y\, D^* \to \pi  D$, and the $f_0(980)$ and $\sigma$ in $Y\, J/\psi \to \pi \pi$, but the induced $s$-channel singularities are further away from the $\sqrt{s}\sim 3900\mev$  region, and give little contribution to the peak.  
 
\begin{figure}[t]
 \centering
 \includegraphics[width=.4\textwidth]{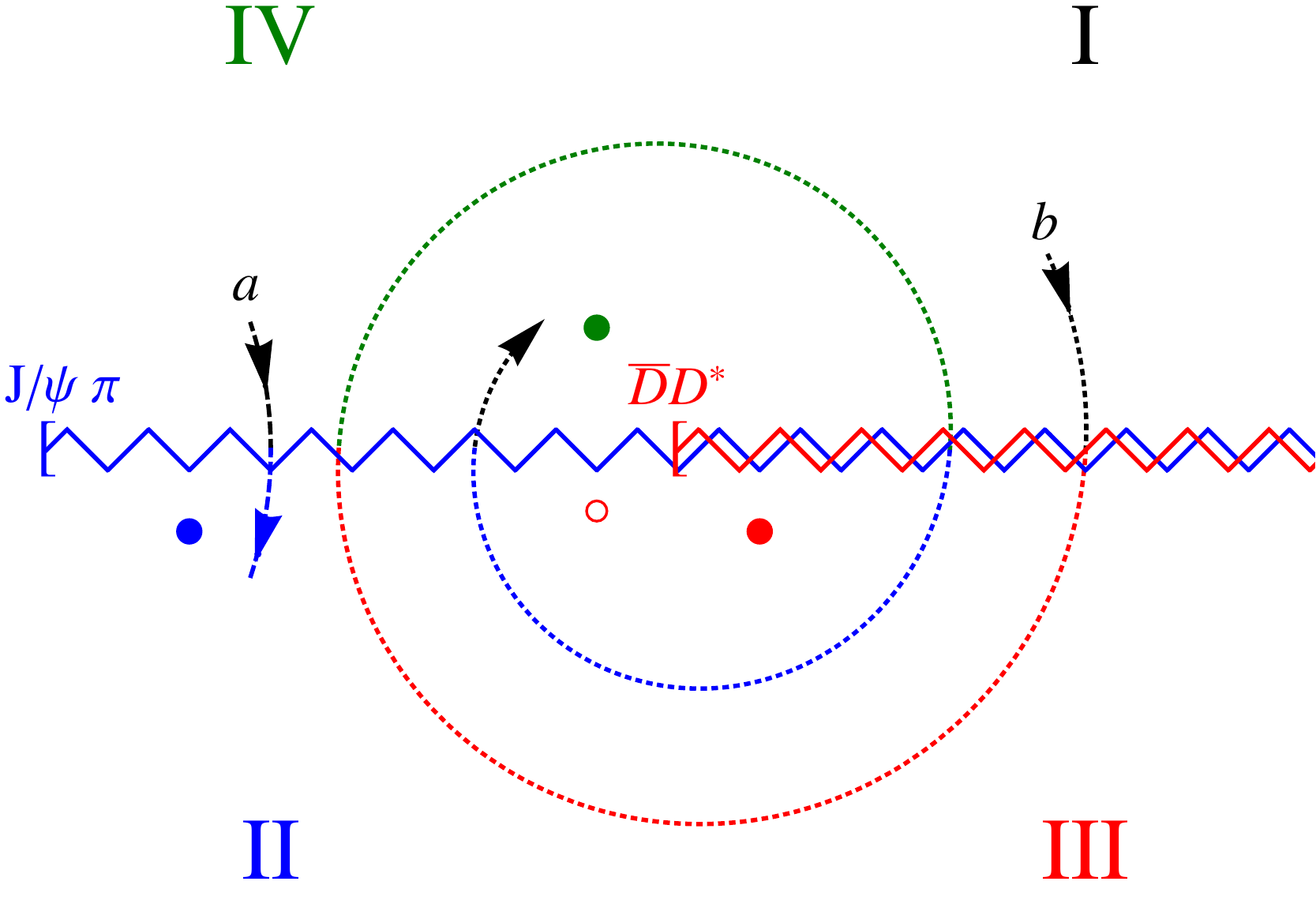}
 \caption{\small Schematic representation of the scattering amplitude $t_{ij}$ as a function of complex $s$. The zig-zag lines represent the unitarity (right-hand) cuts. The physical axis connects to the I sheet right on top of the unitarity cut. Below the (heavier) $\bar D D^*$ threshold, the closest unphysical sheet is II (see path $a$).  A pole on the II sheet below threshold (blue dot), if close enough to the real axis, will produce a Breit-Wigner-like lineshape. Similarly, above the $\bar D D^*$ threshold, the closest unphysical sheet is the III, (see path $b$) and similarly a pole on the III sheet above threshold, if close enough to the real axis, will result  in a Breit-Wigner-like lineshape. On the other hand, poles on the III sheet below threshold (red open circle), or poles in the IV sheet (green disk) are further from the physical region, but can still give rise to a cusp-like peak on the physical axis, if they are close to the $\bar D D^*$ threshold.
}
 \label{fig:pedagogical}
 \end{figure}

If instead the peak is  due  to a pole singularity, the amplitude analysis can provide insights into the (phenomenological) microscopic nature of the $Z_c(3900)$. Consider the schematic plot in \figurename{~\ref{fig:pedagogical}}. The poles related to compact QCD states are expected to become narrower and narrower (\ie approach the real axis of the complex $s$-plane) if the coupling to the open channels is made weaker and weaker (for example, in the large $N_c$ limit~\cite{Weinberg:2013cfa,Knecht:2013yqa,Cohen:2014tga,Cohen:2014vta,Maiani:2016hxw}). Thus, they are expected to be  on the sheet closest to the physical axis, the II sheet if below the \DDstar threshold (blue dot in the figure, reached from the physical axis with path $a$), or the III sheet if above (red dot in the figure, reached from the physical axis with path $b$). A bound state generated by inter-hadron forces would also migrate to the real axis upon switching off of the coupling to the lighter channel, and it is likely to lay on the II sheet as well.
On the other hand, poles on the IV sheet (green dot)  are too far from the physical axis, and would likely stay on the unphysical sheet~\cite{Frazer:1964zz}. The latter case can thus be interpreted as a virtual state, {\it i.e.}  meson-meson configuration for which the attractive interaction is not strong enough to bind the constituents, but nevertheless provides an enhancement in the scattering amplitude, with a typical cusp-like shape.
 
The information about the angular distributions is scarce. The only published plots confirm that the $Y \to \pi Z_c (\to \bar D D^*)$ decay is dominated by the $S$-wave~\cite{Ablikim:2013xfr,Ablikim:2015swa}. In absence of this, there is no point of considering spin. Thus we treat all particles as spinless interacting in the $S$-wave,  at the same time we use physical masses and widths. In doing so, the constraint on the $Y \to \pi Z_c$ angular distribution is automatically fulfilled. One may ask if this approximation is valid for the $D_1$ meson as well, which is known to decay into $D^*\pi$ in $D$-wave~\cite{Aubert:2008zc}. However, the $D$-wave barrier factor (function of $t$) does not affect the $s$-channel projection we are interested in, and can be effectively absorbed in the coupling. Moreover, the helicity distribution ($s$ dependence) turns out to be rather flat if all the  external spins are included, and if the $Y \to \bar D D_1$ decay is dominated by the $S$-wave (the weak sensitivity to this distribution was already commmented in~\cite{Cleven:2013mka}). In this respect, the spinless $S$-wave approximation gives a more realistic description with respect to a spinless $D$-wave treatment, of the $D_1$.  This choice reduces the number of free parameters, and only turns out in a poorer description of the reflected peak in the  $\jpsi \,\pi$ channel at $\sim 3.45\gev$.   

We denote by $f_i(s,t,u)$ the scalar amplitudes for the two reactions shown in \figurename{~\ref{fig:diagrams}}, with 
  $i=1$ referring to $Y\pi \to \bar D D^*$ and $i=2$ to $Y\pi \to \jpsi \pi$. 
 These are given by sums over a finite number of isobar amplitudes in the various channels~\cite{Khuri:1960zz},
\begin{equation}
f_i(s,t,u) = 16\pi \sum_{l=0}^{L_\text{max}} (2l + 1) 
 \left(a^{(s)}_{l,i}(s)P_l(z_s) +a^{(t)}_{l,i}(t)P_l(z_t)  +a^{(u)}_{l,i}(u)P_l(z_u)  \right)
\end{equation}
with $z_x$ being the cosine of the scattering angle in the center of mass frame of the $x=s,t,u$ channel. 
We consider all the exchanges to happen in $S$-wave, the higher waves being kinematically suppressed, $a_{l,i}^{(x)} = 0$, for $l>0$. The $s$-channel partial wave amplitudes are given by 
  \begin{subequations}
\begin{align}
f_{0,i}(s) &= \frac{1}{32\pi} \int_{-1}^1 d z_s\, f_i\left(s, t(s, z_s), u(s, z_s)\right) = a_{0,i}^{(s)} + \frac{1}{32\pi} \int_{-1}^1 d z_s \left(a^{(t)}_{0,i}(t)  +a^{(u)}_{0,i}(u) \right) 
\equiv a_{0,i}^{(s)} + b_{0,i}(s)\\
f_{l,i}(s) &= \frac{1}{32\pi} \int_{-1}^1 d z_s \, P_l(z_s) \left(a^{(t)}_{0,i}(t)  +a^{(u)}_{0,i}(u) \right)\equiv b_{l,i}(s)\quad\text{for $l>0$}.
\end{align}
 \end{subequations}
By construction, the isobars $a_{0,i}$ contain right hand singularities only, whereas the projections of the crossed channels isobars
 induce left hand singularities in the  $b_{0,i}$ amplitudes. 
Unitarity determines the discontinuity $\Delta f_{l,i}(s) = \tfrac{1}{2i}\left(f_{l,i}(s + i\epsilon) - f_{l,i}(s - i\epsilon)\right)$ across the right hand cut,
 \begin{subequations}
\begin{align} 
\Delta f_{0,i}(s) &= \sum_j t^*_{ij}(s) \,\rho_j(s) \,f_{0,j}(s) \label{eq:unitarity} \\
\Delta f_{l,i}(s) &= 0\quad\text{for $l>0$}
\end{align}
 \end{subequations}
with $t_{ij}$ the $2 \times 2$ $S$-wave scattering matrix, and $\rho_j$ the phase space in the $j$ channel, \ie $\rho_j(s) = \lambda^{1/2}\left(s, m^2_{j,1}, m^2_{j,2}\right)/s$. The solution to Eq.~\eqref{eq:unitarity} is given by  the well known 
Omn\`es representation~\cite{Szczepaniak:2015hya}, 
\begin{equation} 
f_{0,i}(s) = b_{0,i}(s)  + \sum_j t_{ij}(s)\frac{1}{\pi} \int_{s_j}^\infty  ds^\prime  \frac{\rho_j(s^\prime) b_{0,j}(s^\prime) }{s^\prime - s}\label{eq:fsi},
\end{equation}
with $s_j$ the threshold of channel $j$. We ignored possible contributions from left hand singularities in the scattering matrix. We subtract the integral once to improve its convergence, and to take into account any other short-range exchange.

The original projections $b_{l,i}(s)$ are thus modified by an additional term describing the final state interactions. If this happens only for a finite number of partial waves (only $S$-waves in this model, Eq.~\eqref{eq:fsi}), the partial wave series can be summed back and it simply reconstructs the original isobars in the crossed channel,  
\begin{equation} 
f_i(s,t,u) = 16\pi\,\left[a^{(t)}_{0,i}(t) +a^{(u)}_{0,i}(u) + \sum_j t_{ij}(s)\left(c_j + \frac{s}{\pi}  \int_{s_j}^\infty  ds^\prime  \frac{\rho_j(s^\prime) b_{0,j}(s^\prime) }{s^\prime\left(s^\prime - s\right)}\right)\right], \label{eq:sol} 
\end{equation}
where the subtraction constants $c_j$ are explicitly shown.
We do not expect the fits to the Dalitz plot projections to be sensitive to details of the lineshapes in the crossed channel, so we parametrize the
 isobar amplitudes as simple Breit-Wigners,
 \begin{equation}
16\pi \, a^{(x)}_{0,i}(x) = \sum_r \frac{g_r}{m_r^2 - x - i m_r \Gamma_r }  \equiv \sum_r  BW\!\left(x, r\right),
 \end{equation}
with $x = t,u$. With these, we define the  dispersed projections, {\it cf.} Eq.~\eqref{eq:sol} by 
  \begin{equation}
H\!\left(s, r\right) \equiv \frac{s}{\pi} \int_{s_r}^\infty  ds^\prime  \frac{\rho_r(s^\prime)  }{s^\prime \left(s^\prime - s\right)} \int_{-1}^1 \frac{dz}{2} BW\!\left(x(s^\prime,z), r\right)
 \end{equation}
  with $r = D_0,\, D_1,\, f_0, \,\sigma$ referring to the various cross channel exchanges that we take into account. The thresholds $s_r$ and the phase space $\rho_r$ are related to the channel the exchanged resonance appears in, namely channel 1 for $D_0,\,D_1$ 
  and channel 2 for $f_0$, $\sigma$. The final expressions for the  amplitudes are  given by 
 \begin{align}
 f_1(s,t,u) &= BW\!\left(t, D_1\right) +  BW\!\left(u, D_0\right)   + t_{11}(s) \left[c_1 + H(s, D_1) + H(s,D_0)\right] + t_{12}(s) \left[c_2 + H(s, f_0) + H(s,\sigma)\right]\label{eq:ampl1},
 \end{align}
and
 \begin{align}
 f_2(s,t,u) &= BW\!\left(t, f_0\right) +  BW\!\left(t, \sigma\right)  + t_{21}(s) \left[c_1 + H(s, D_1) + H(s,D_0)\right] + t_{22}(s) \left[c_2 + H(s, f_0) + H(s,\sigma)\right]  \nonumber\\ &\quad +\left(s \leftrightarrow u\right)\label{eq:ampl2}.
  \end{align}
and the expression for the Dalitz projections being
\begin{equation}
\frac{d\Gamma_i}{d\!\sqrt{s}} \propto \sqrt{s} \int_{t_\text{min}(s)}^{t_\text{max}(s)} dt\,\left|f_i\left(s,t,u(s,t)\right)\right|^2,
\end{equation}
and similarly for the projections in the $\sqrt{t}$ or $\sqrt{u}$ variables.
Until now, we have not given any detail on the nature of the $t_{ij}$ scattering matrix. We use a $K$ matrix parametrization, $t_{ij}(s) = \left(K^{-1} - i \rho(s)\right)^{-1}_{ij}$
with $\rho_{ij} = \rho_i \delta_{ij}$. This parametrization contains spurious left hand cuts, which we remove by approximating~\cite{Fernandez-Ramirez:2015tfa}
\begin{equation}
\rho_{i} = \frac{\sqrt{\left(s - (m_{1,i} + m_{2,i})^2\right)\left(s - (m_{1,i} - m_{2,i})^2\right)}}{s} \simeq \sqrt{s - (m_{1,i} + m_{2,i})^2} \frac{2 \sqrt{m_{1,i} m_{2,i}}}{\left(m_{1,i} + m_{2,i}\right)^2}.
\end{equation}
Alternatively, we have considered a Chew-Mandelstam phase space, but this choice has very little impact on the fits, and we will not discuss it any further. We consider four different scenarios:
\begin{enumerate}
\item \scenIII: In this case we consider the parametrization which is as close as possible to the one used in the original experimental analyses, even if it violates unitarity. To wit, for the $K$ matrix we use the Flatt\'e parametrization, \ie $K_{ij} = g_i g_j / (M^2 - s)$. This choice produces poles in the closest sheet to the real axis, \ie the III sheet above the \DDstar threshold, or the II sheet below threshold. The former case might be interpreted as a genuine 
 QCD state, the latter could be a QCD state of a hadron molecule. We artificially remove the influence of triangle singularities by imposing $H = 0$, thus breaking unitarity.
\item \scenIIItr: The $K$ matrix is as in case ``\scenIII'',  but we reinstate the correct value for $H$, which gives rise to a triangle singularity close to the physical region. That is, the $S$-waves in the $s$-channel near the physical region, can 
 have both the resonance pole and the logarithmic branch point from the triangle singularity. 
\item \scenIV: In this case we choose for $K$ a symmetric constant matrix. This choice can produce poles in the IV Riemann sheet that can be interpreted as virtual states with respect to the heavier \DDstar channel. This would be more likely due to hadron-hadron interactions. 
\item \scentr: The $K$ matrix is as in case ``\scenIV' except that we force the possible pole in $t$ to be far from the $\jpsi \, \pi$ threshold. We do this by imposing a penalty on the $\chi^2$ which linearly decrease with the distance of the pole from the point $s_0 = 15\gev^2$,  which corresponds to the position of the peak, and vanishes outside the disk of radius $10\gev^2$. With this model we can assess whether the triangle singularity alone is able to generate the observed structure.
\end{enumerate}

Similar analyses have been perfomed in~\cite{Albaladejo:2015lob,Albaladejo:2016jsg}, considering scenarios comparable with \scenIIItr and \scenIV, although the parametrizations employed are different.

 \begin{figure}[p]
 \centering
 \begin{subfigure}[t]{.32\columnwidth}
\includegraphics[width=\textwidth]{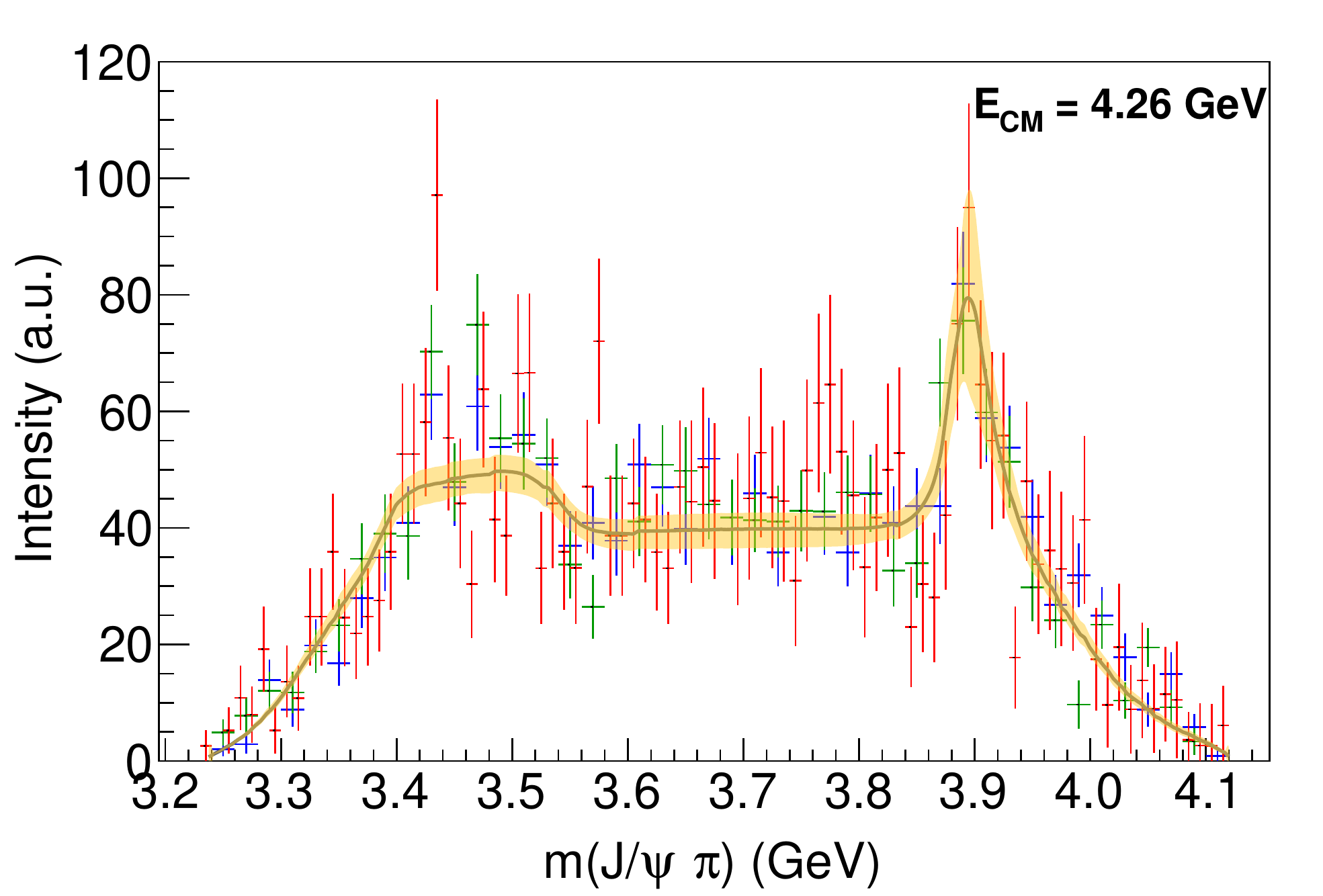}
\caption{\small ~}
\end{subfigure}
 \begin{subfigure}[t]{.32\columnwidth}
\includegraphics[width=\textwidth]{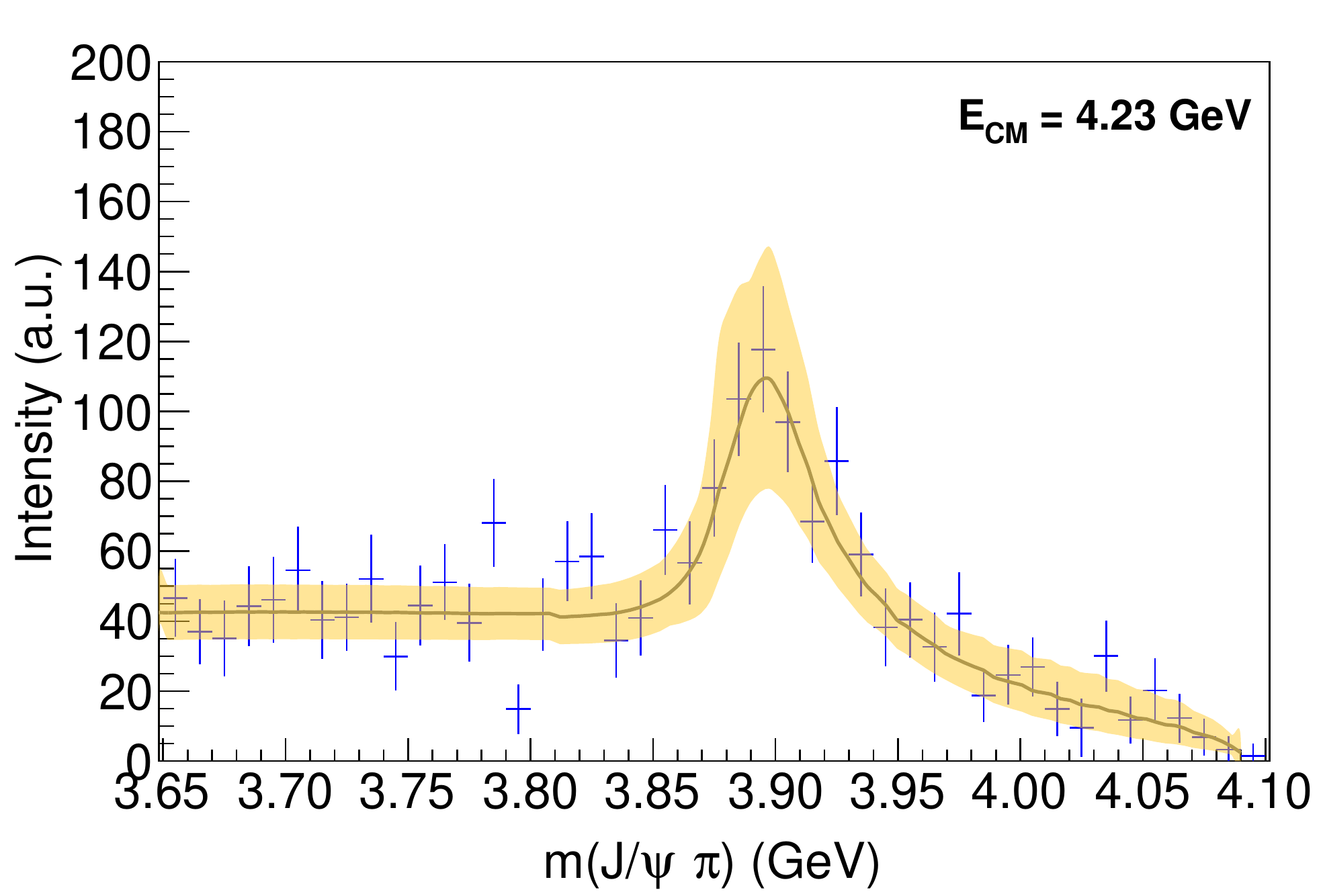}
\caption{\small ~}
\end{subfigure}
 \begin{subfigure}[t]{.32\columnwidth}
\includegraphics[width=\textwidth]{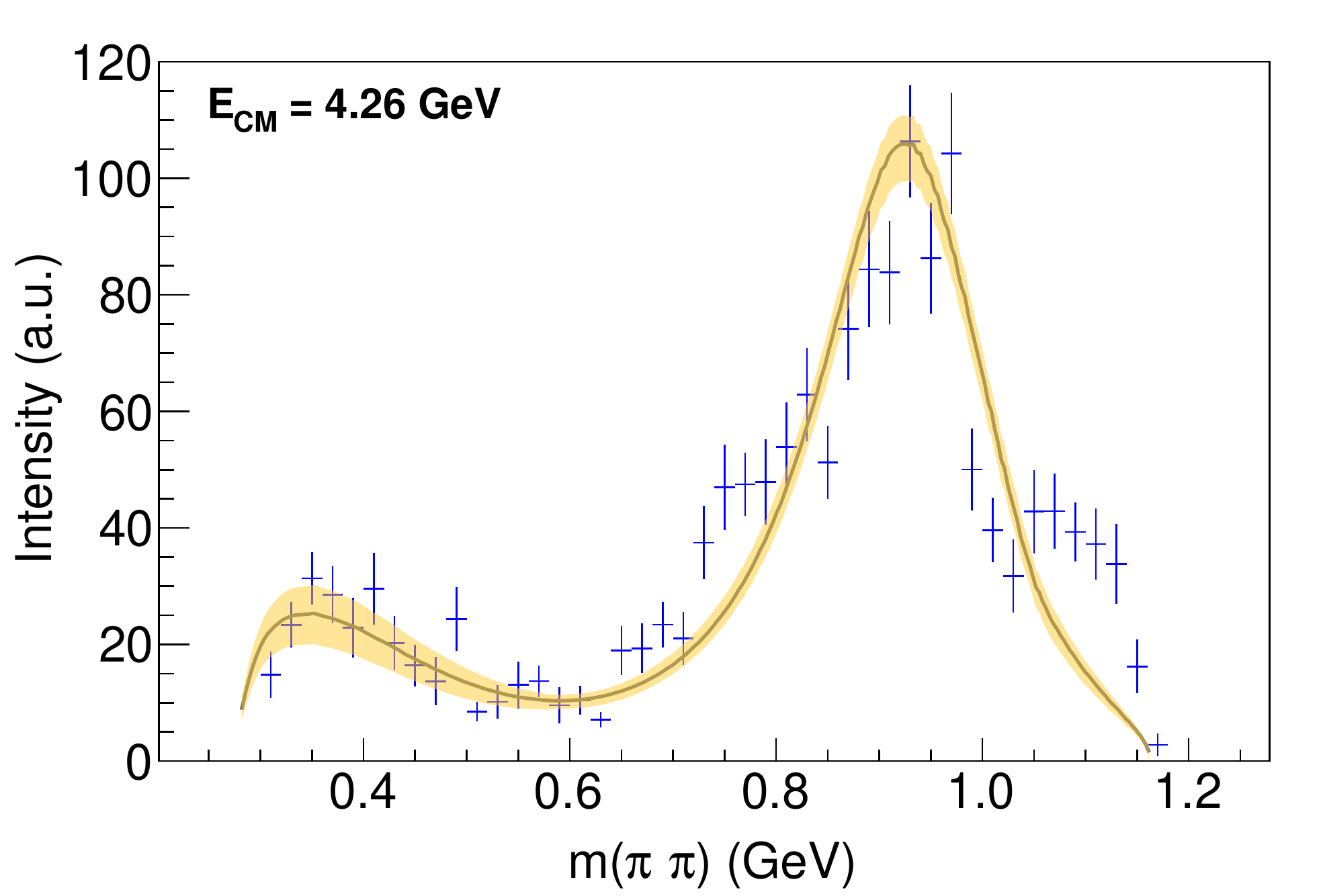}
\caption{\small ~}
\end{subfigure}\\
 \begin{subfigure}[t]{.32\columnwidth}
\includegraphics[width=\textwidth]{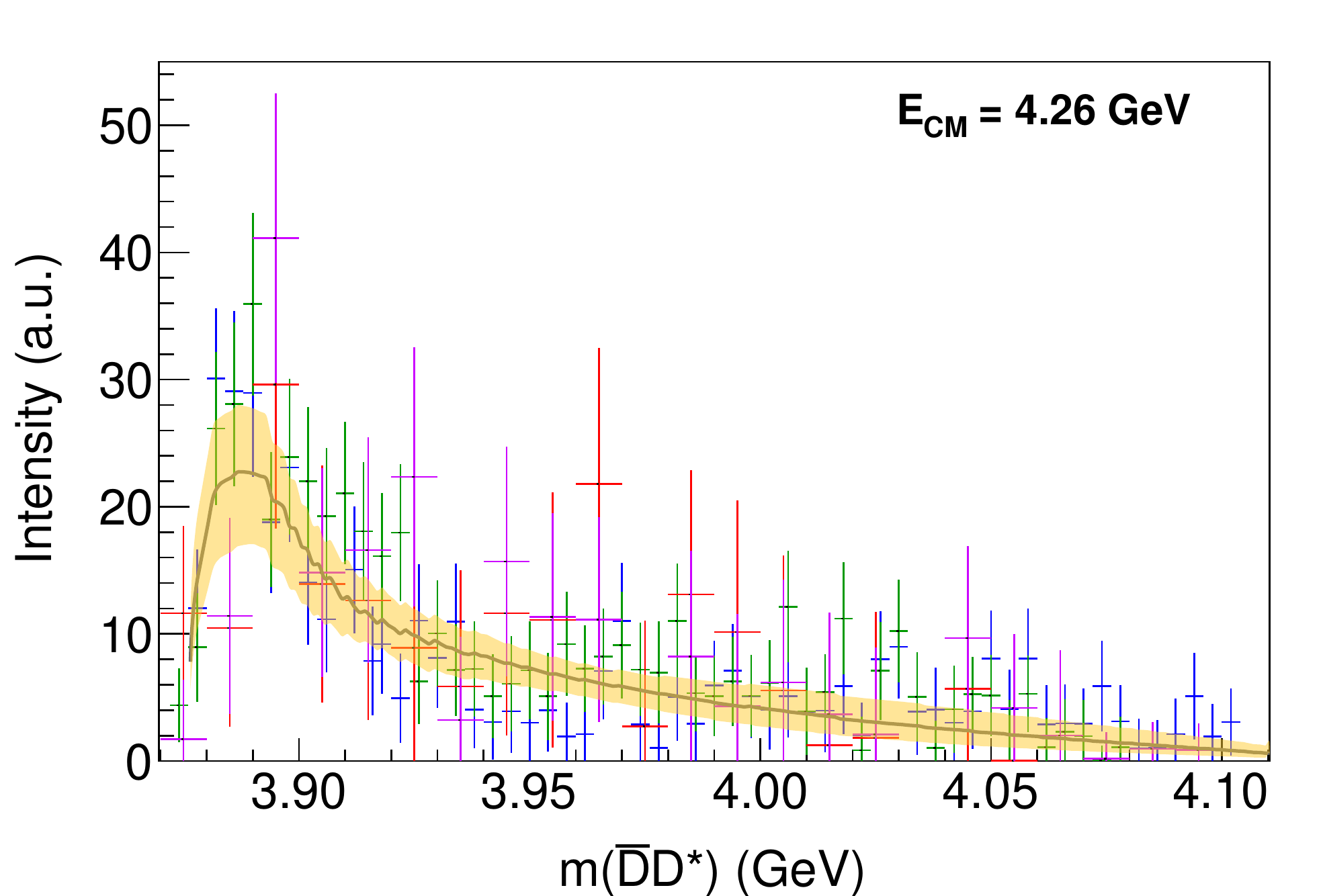}
\caption{\small ~}
\end{subfigure}
 \begin{subfigure}[t]{.32\columnwidth}
\includegraphics[width=\textwidth]{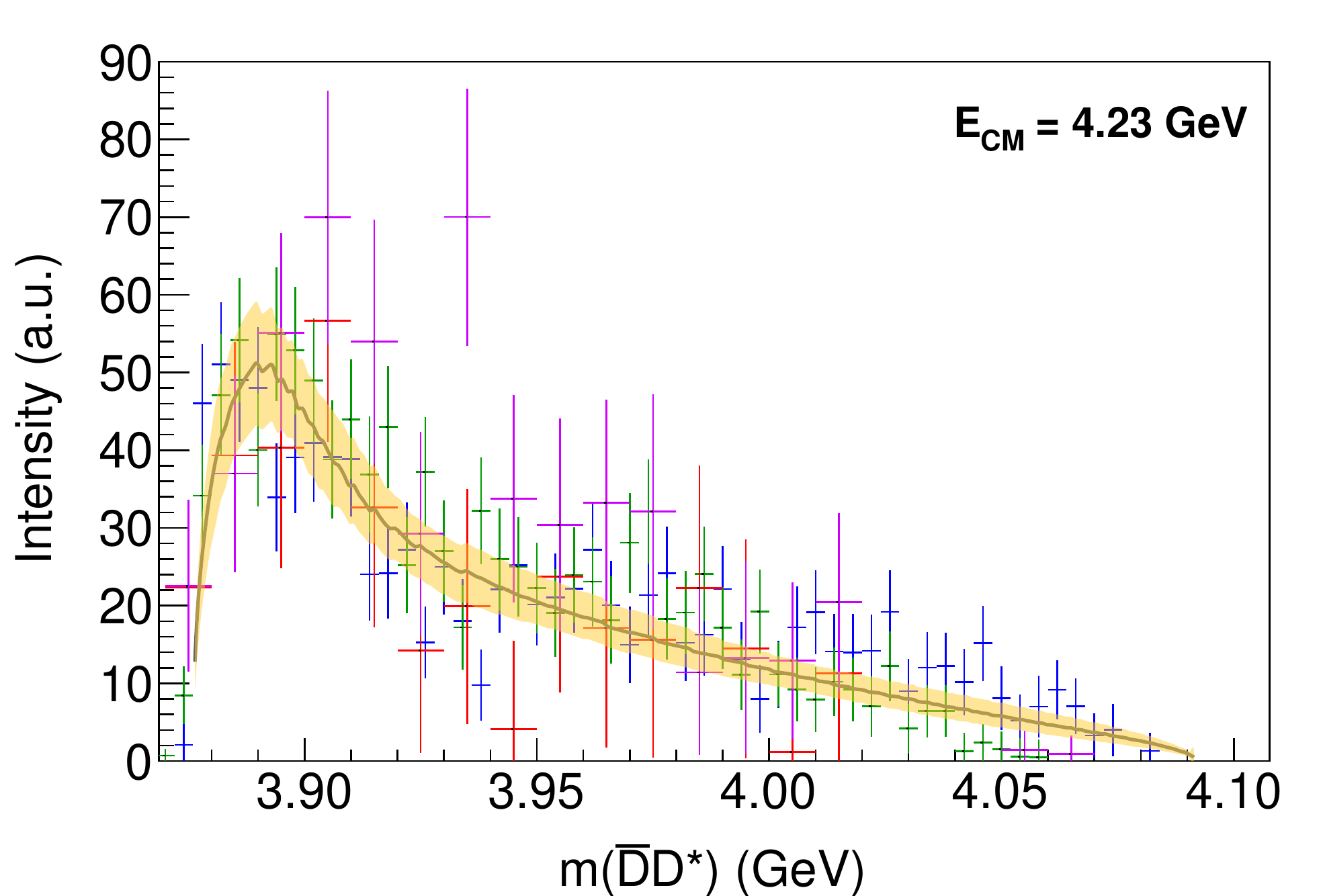}
\caption{\small ~}
\end{subfigure}
 \caption{Result of the fit for the  scenario \scenIII (Flatt\'e $K$-matrix, without triangle singularity). The grey line and the yellow band show the fit result with the relative $1\sigma$ error. (a) $\jpsi \pi$ projection of the $Y(4260)\to \jpsi \pi\pi$ reaction at $E_\text{CM} = 4.26\gev$. Green (blue) points are the $\jpsi \pi^+$  ($\jpsi \pi^-$) data~\cite{Ablikim:2013mio}; red points are the $\jpsi \pi^0$ data \cite{Ablikim:2015tbp}, rescaled as described in the text. As expected, the fit does not reproduce the peaking structure at $3.45$~GeV, which is the reflection of the peak at the right, and would require to take spins properly into account. (b) $\jpsi \pi^0$ projection of the $Y(4260)\to \jpsi \pi^0\pi^0$ at $E_\text{CM} = 4.23\gev$~\cite{Ablikim:2015tbp}, rescaled as described in the text. (c) $\pi \pi$ projection of the $Y(4260)\to \jpsi \pi\pi$ at $E_\text{CM} = 4.26\gev$~\cite{Ablikim:2013mio}; the points at $m_{\pi\pi} > 1$~GeV are not described by the simple two resonances model, and are excluded from the fit.
 (d) $\bar DD^*$ projection of the $Y(4260) \to \bar D D^* \pi$  reaction at $E_\text{CM} = 4.26\gev$. Green (blue) points are the $D^- D^{*0}$ ($D^0 D^{*-}$) data~\cite{Ablikim:2015swa}; red (purple) points are the $\bar D^0 D^{*0}$ ($D^+ D^{*-}$) data~\cite{Ablikim:2015gda}, rescaled and background-subtracted as explained in the text.  (e) same as (d), but for $E_\text{CM} = 4.23 \gev$. The errors shown are statistical only.}
 \label{fig:scenIII}
\vspace{1cm}
 \begin{subfigure}[t]{.32\columnwidth}
\includegraphics[width=\textwidth]{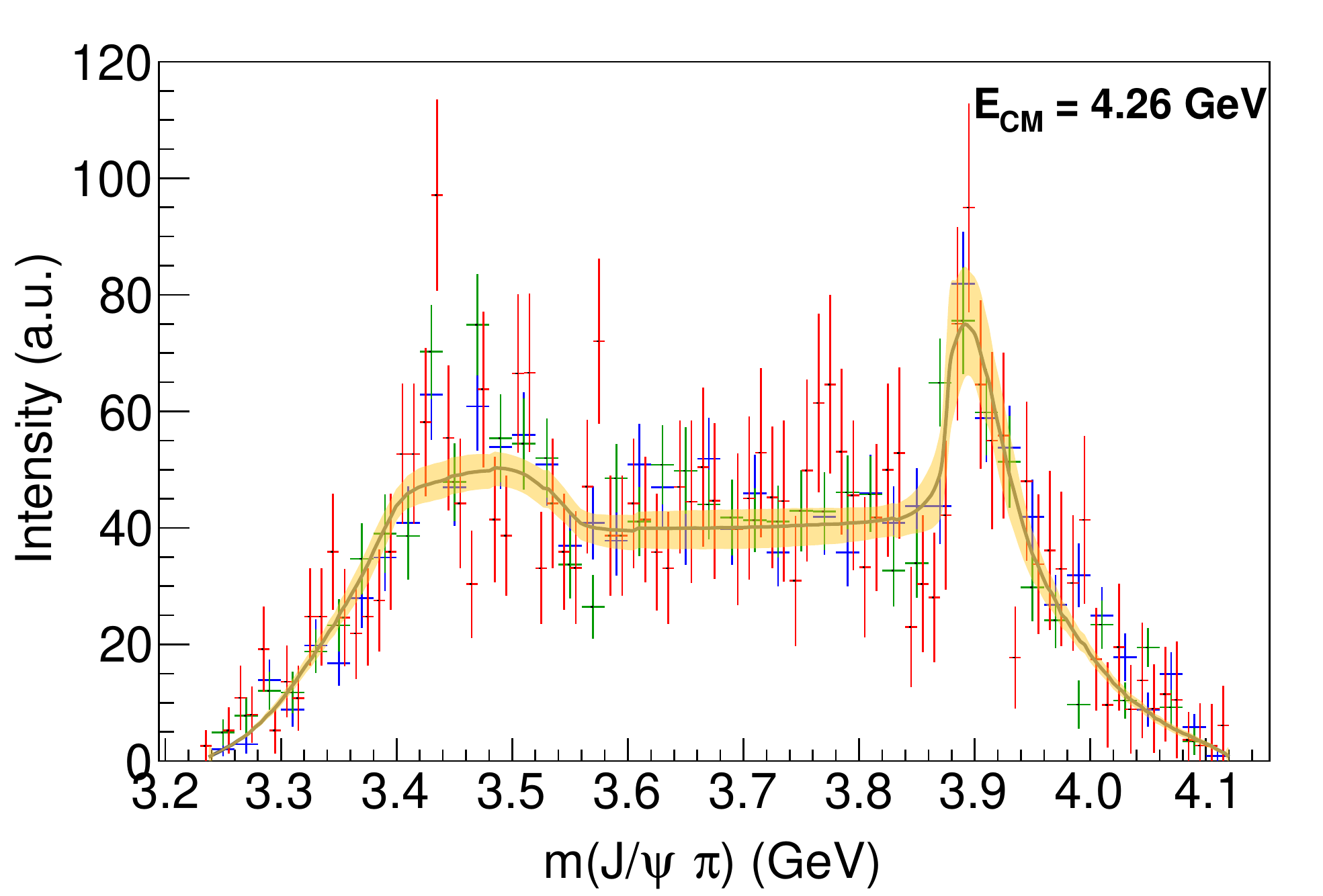}
\caption{\small ~}
\end{subfigure}
 \begin{subfigure}[t]{.32\columnwidth}
\includegraphics[width=\textwidth]{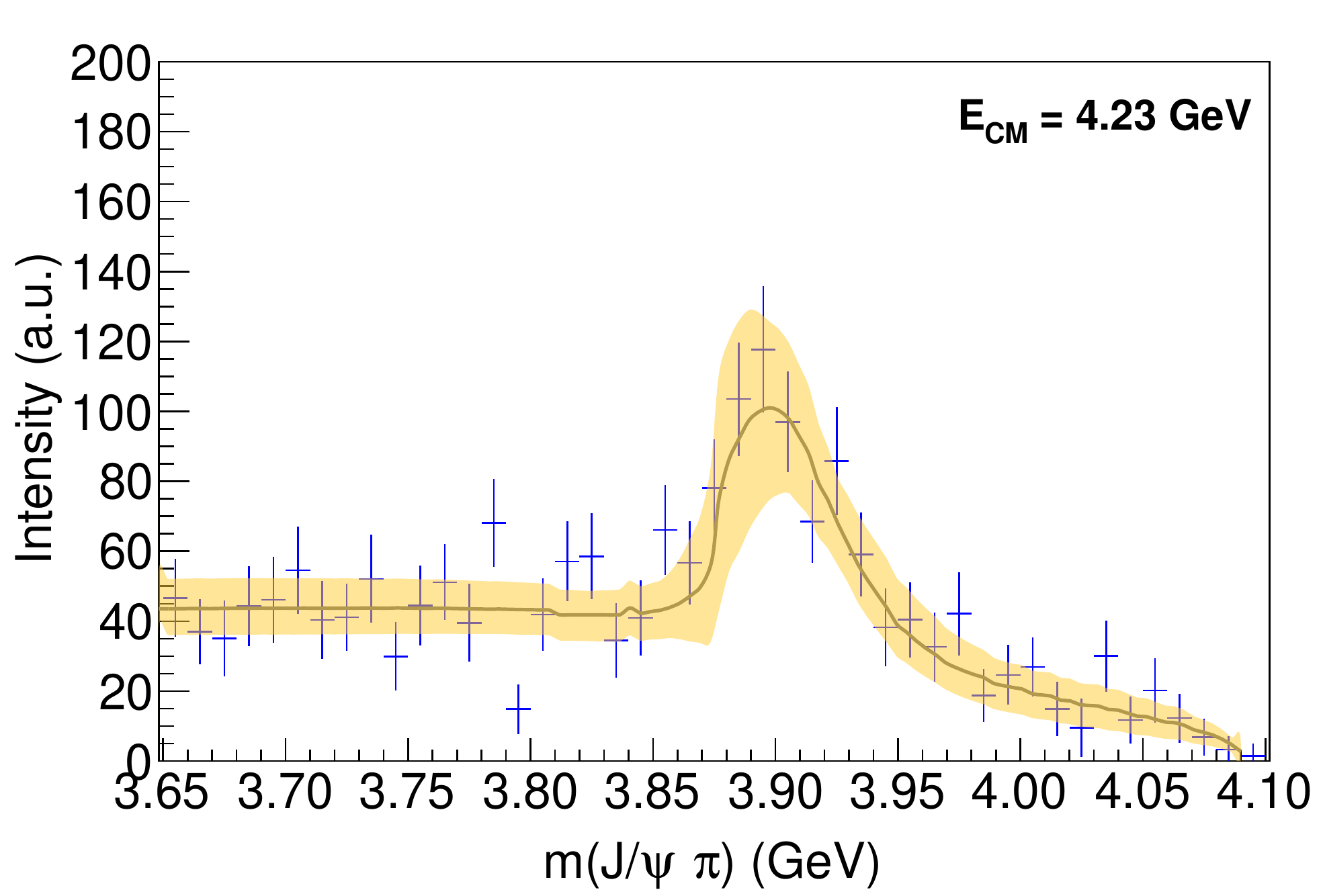}
\caption{\small ~}
\end{subfigure}
 \begin{subfigure}[t]{.32\columnwidth}
\includegraphics[width=\textwidth]{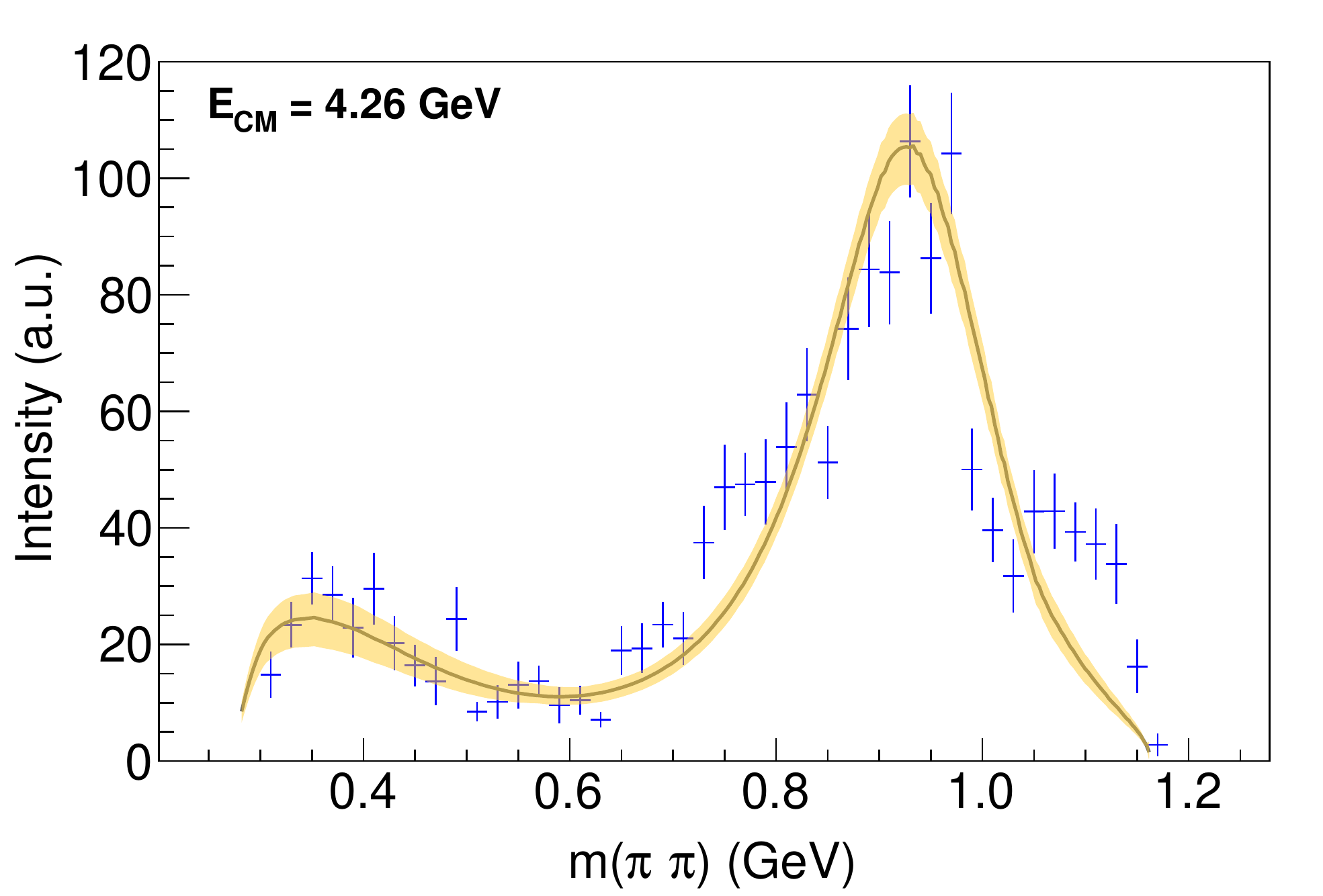}
\caption{\small ~}
\end{subfigure}\\
 \begin{subfigure}[t]{.32\columnwidth}
\includegraphics[width=\textwidth]{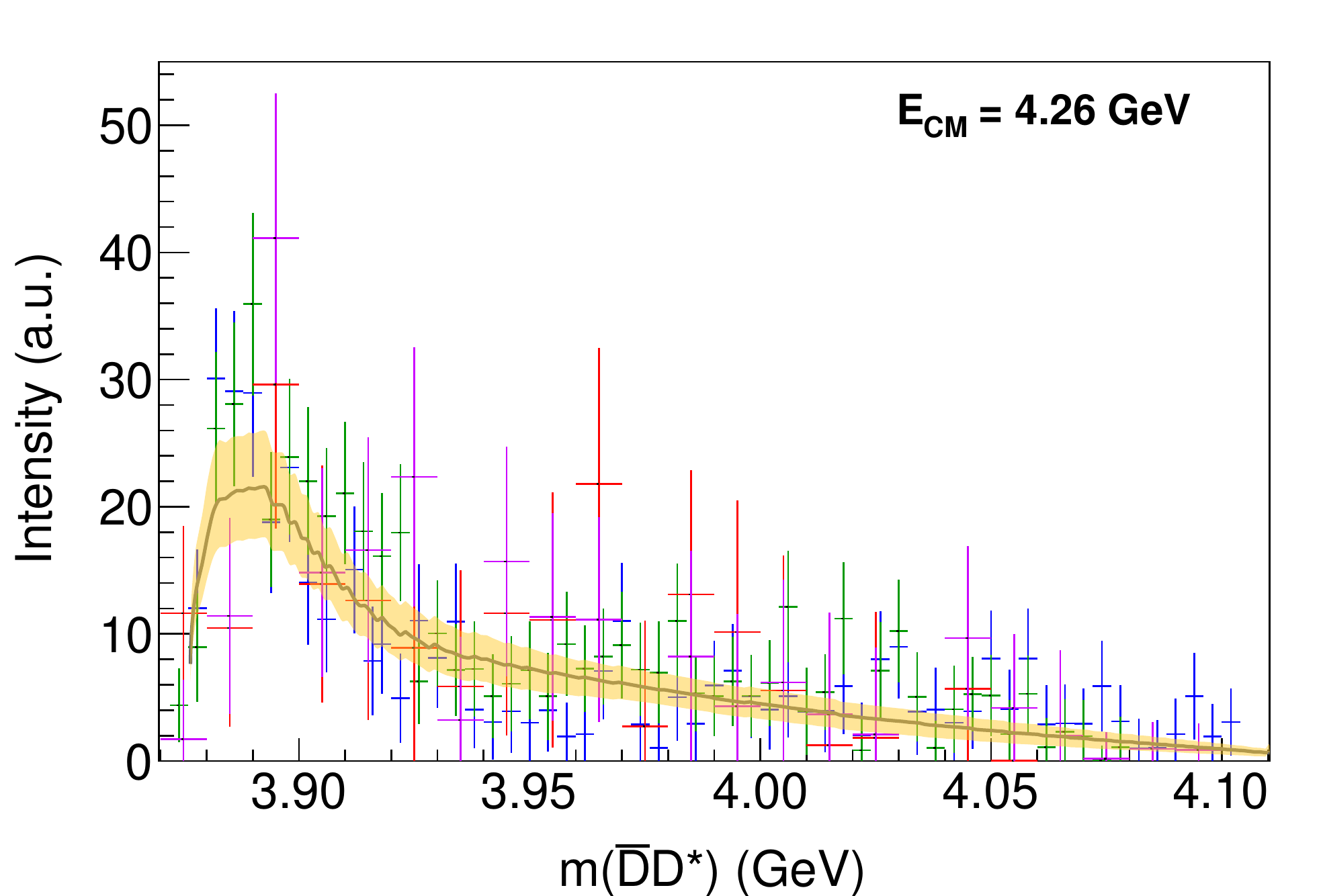}
\caption{\small ~}
\end{subfigure}
 \begin{subfigure}[t]{.32\columnwidth}
\includegraphics[width=\textwidth]{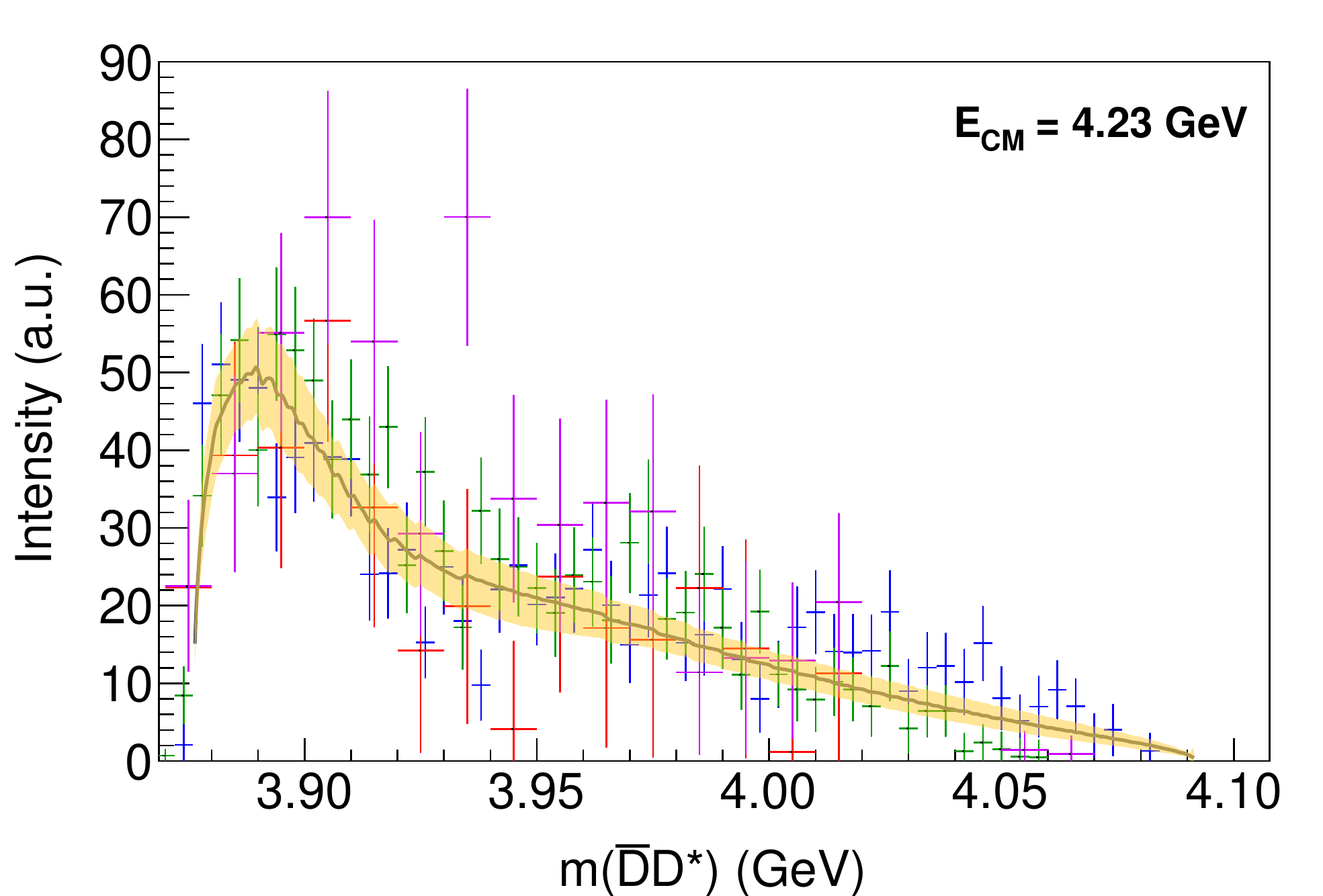}
\caption{\small ~}
\end{subfigure}
 \caption{Result of the fit for the  scenario \scenIIItr (Flatt\'e $K$-matrix, with triangle singularity). The plot legend and the comments on the fit are given in the caption of \figurename{~\ref{fig:scenIII}}. }
 \label{fig:scenIIItr}
 \end{figure}
 
  \begin{figure}[p]
 \centering
 \begin{subfigure}[t]{.32\columnwidth}
\includegraphics[width=\textwidth]{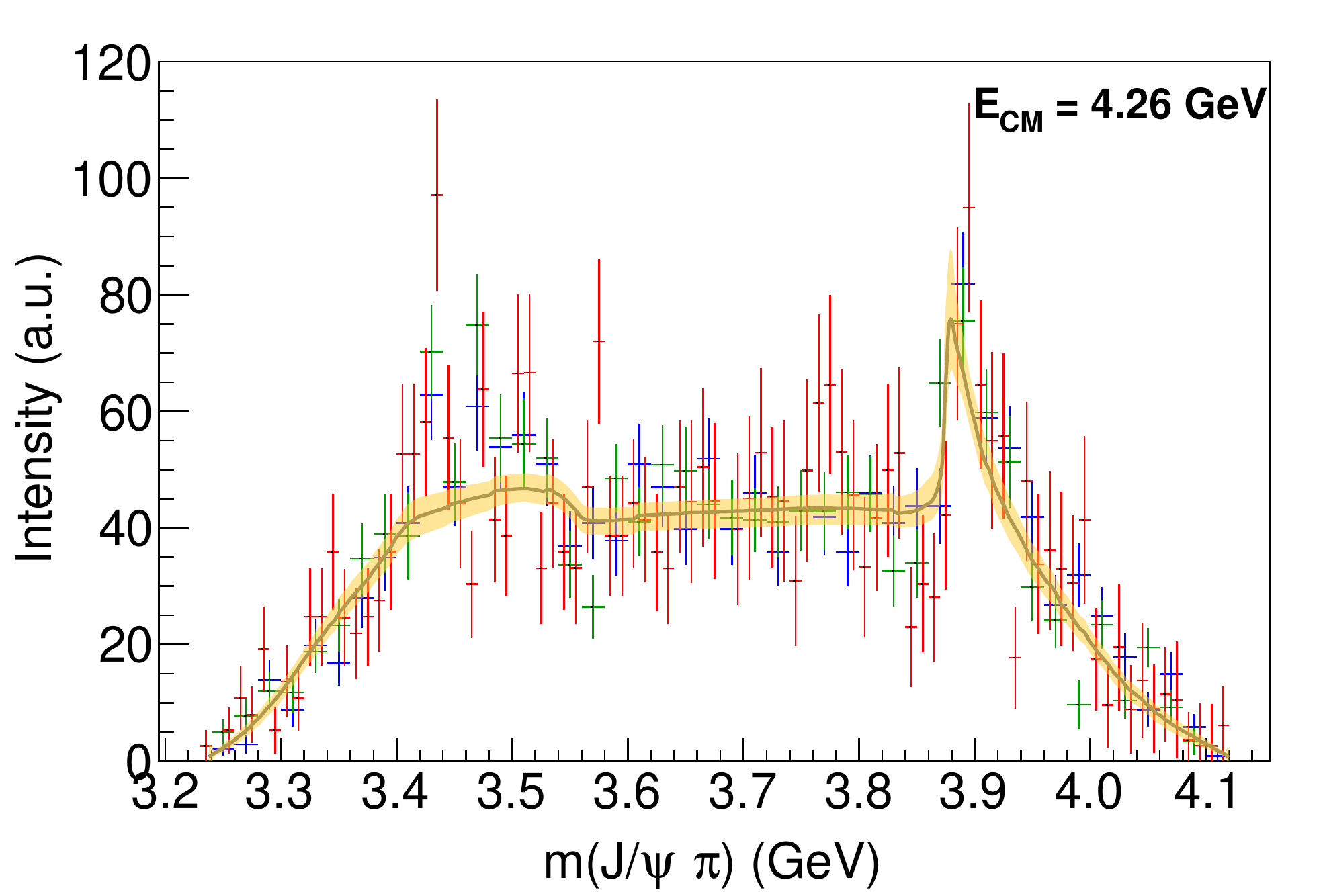}
\caption{\small ~}
\end{subfigure}
 \begin{subfigure}[t]{.32\columnwidth}
\includegraphics[width=\textwidth]{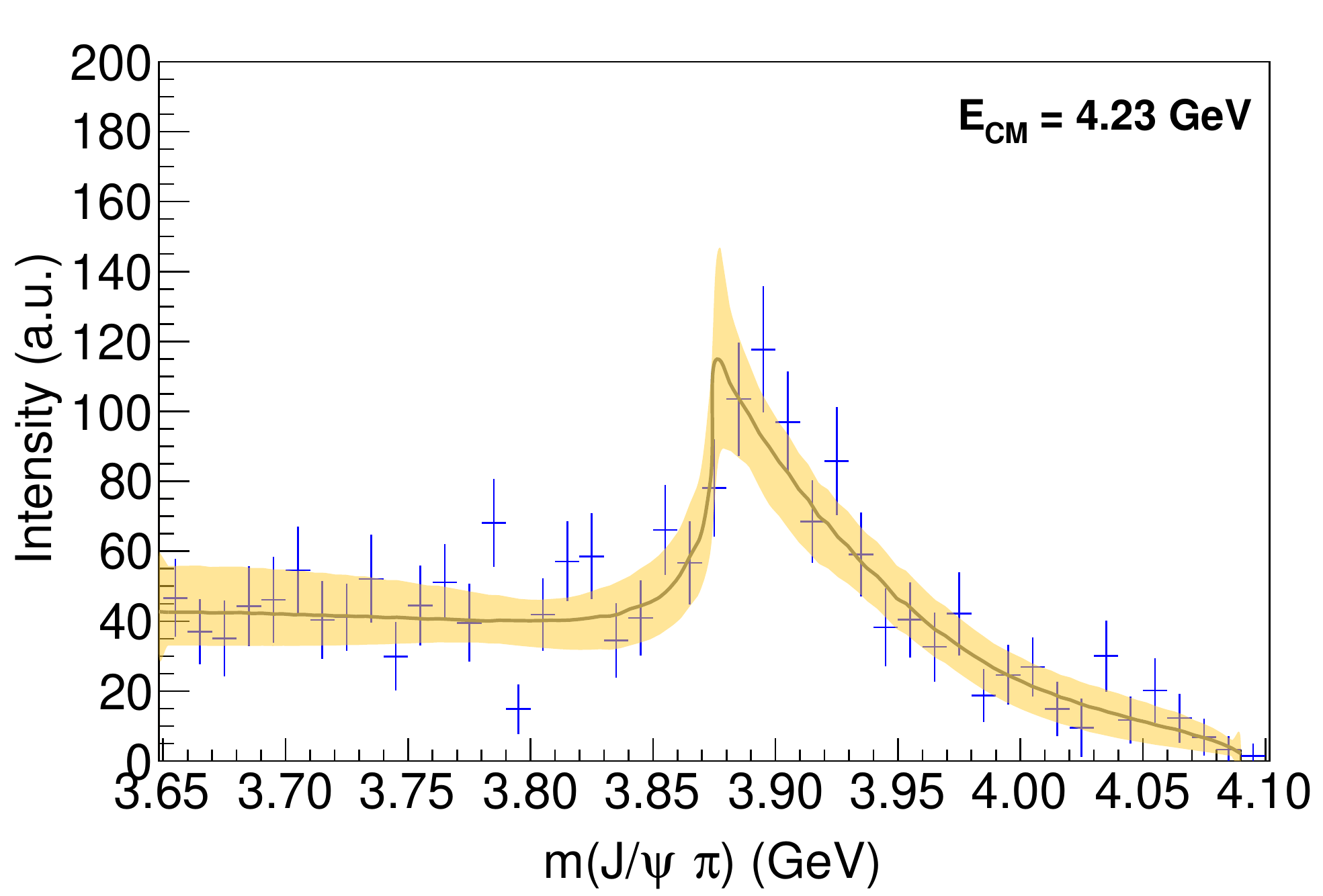}
\caption{\small ~}
\end{subfigure}
 \begin{subfigure}[t]{.32\columnwidth}
\includegraphics[width=\textwidth]{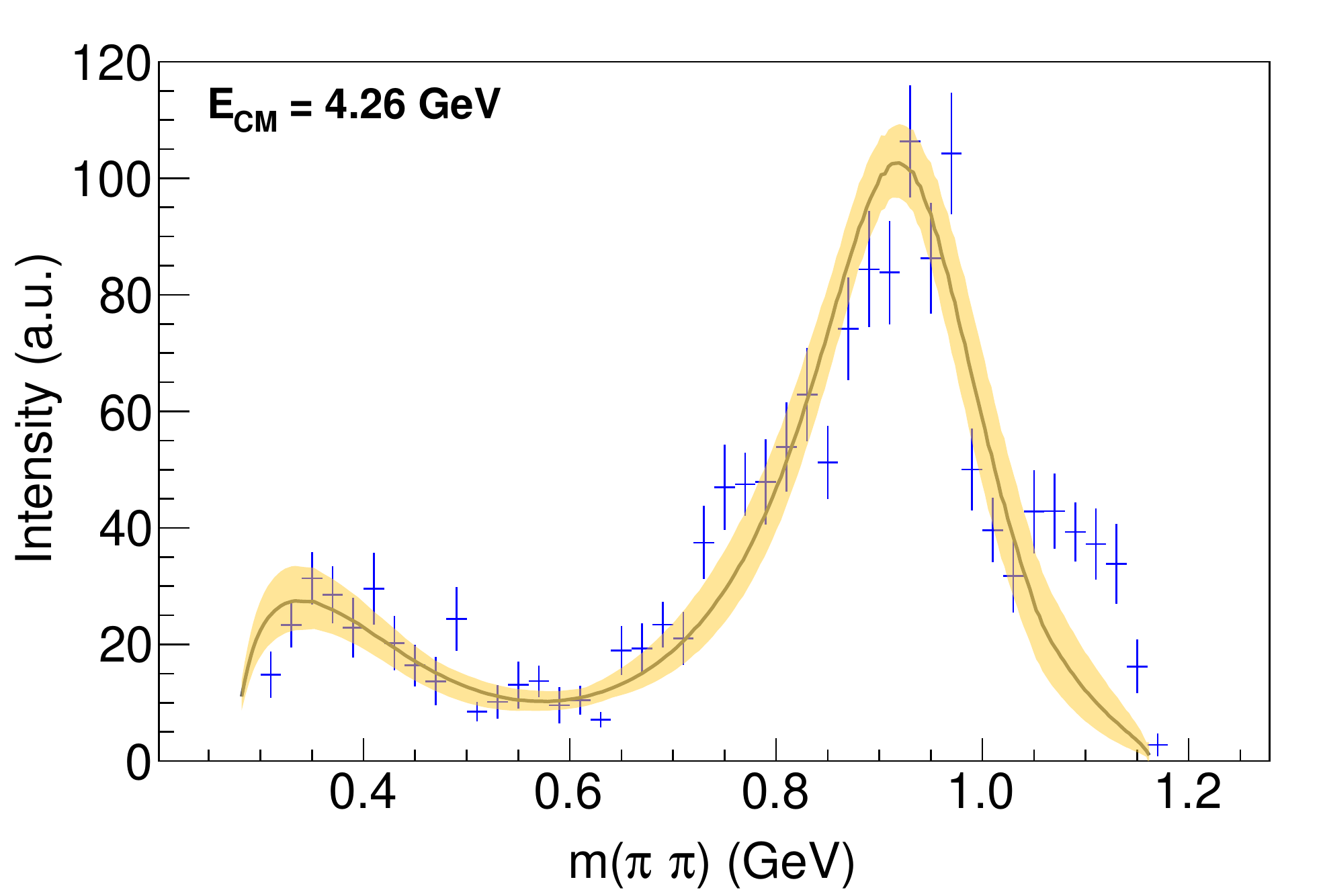}
\caption{\small ~}
\end{subfigure}\\
 \begin{subfigure}[t]{.32\columnwidth}
\includegraphics[width=\textwidth]{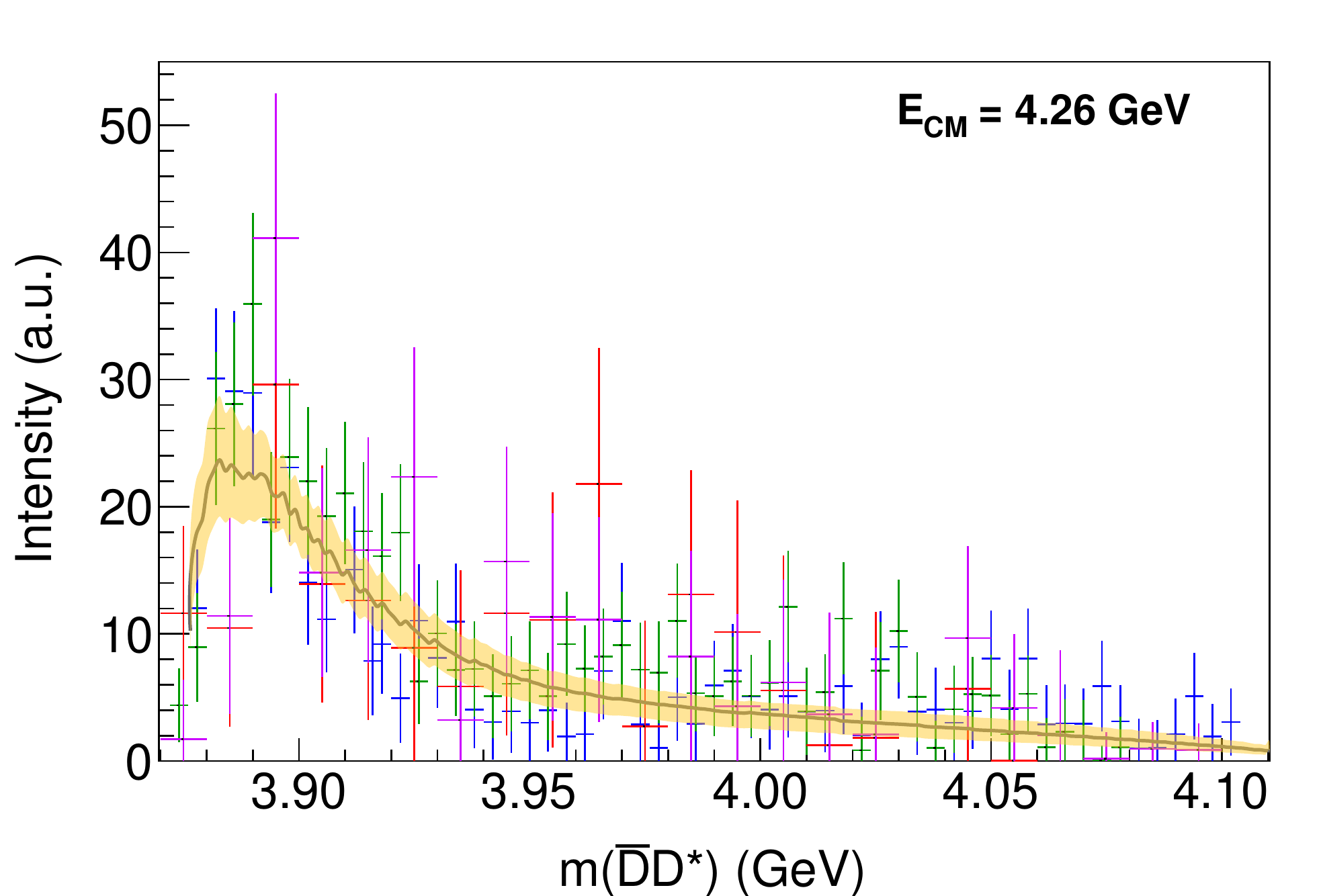}
\caption{\small ~}
\end{subfigure}
 \begin{subfigure}[t]{.32\columnwidth}
\includegraphics[width=\textwidth]{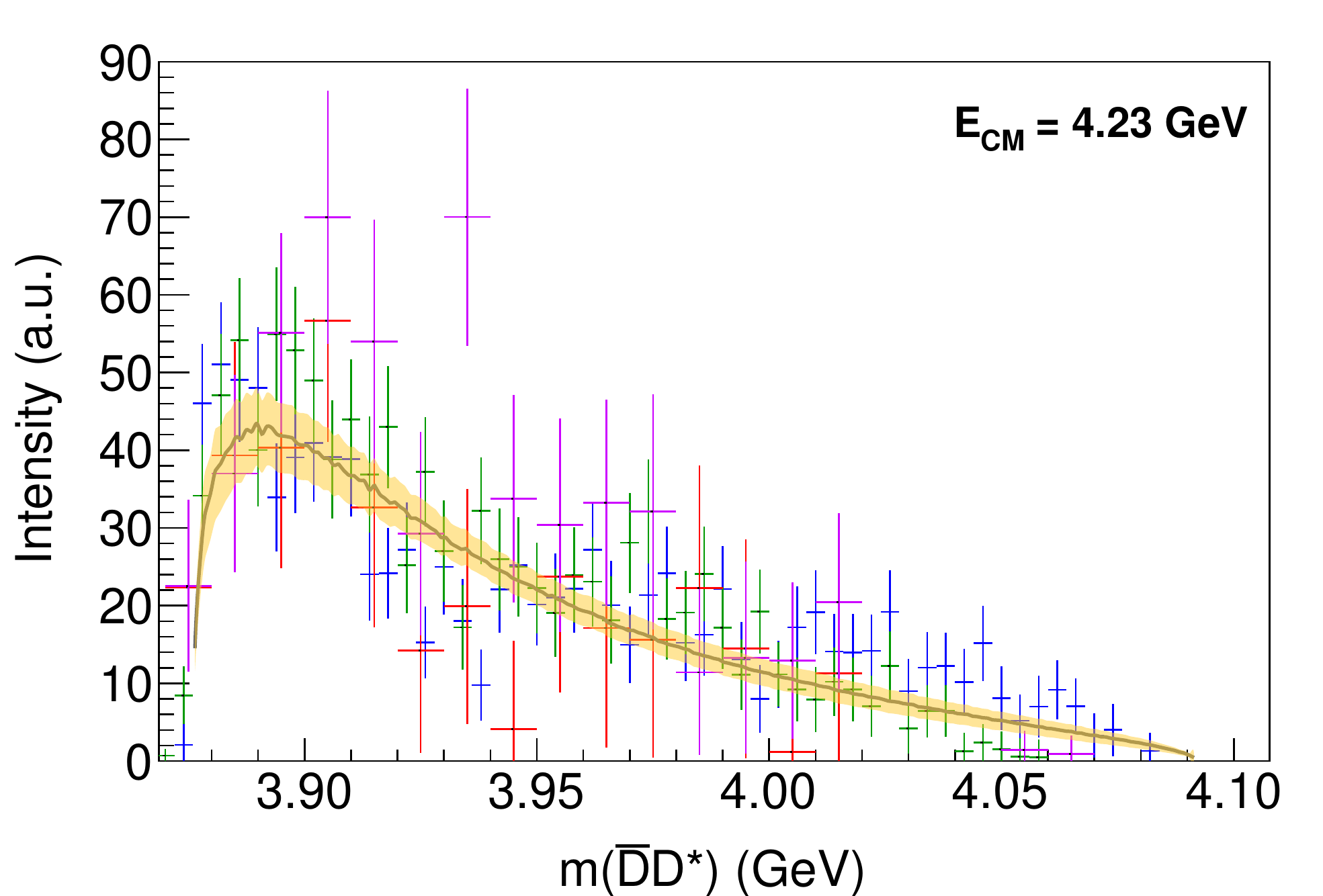}
\caption{\small ~}
\end{subfigure}
 \caption{Result of the fit for the  scenario \scenIV (constant $K$-matrix, with triangle singularity). The plot legend and the comments on the fit are given in the caption of \figurename{~\ref{fig:scenIII}}. }
 \label{fig:scenIV}
\vspace{1cm}
 \begin{subfigure}[t]{.32\columnwidth}
\includegraphics[width=\textwidth]{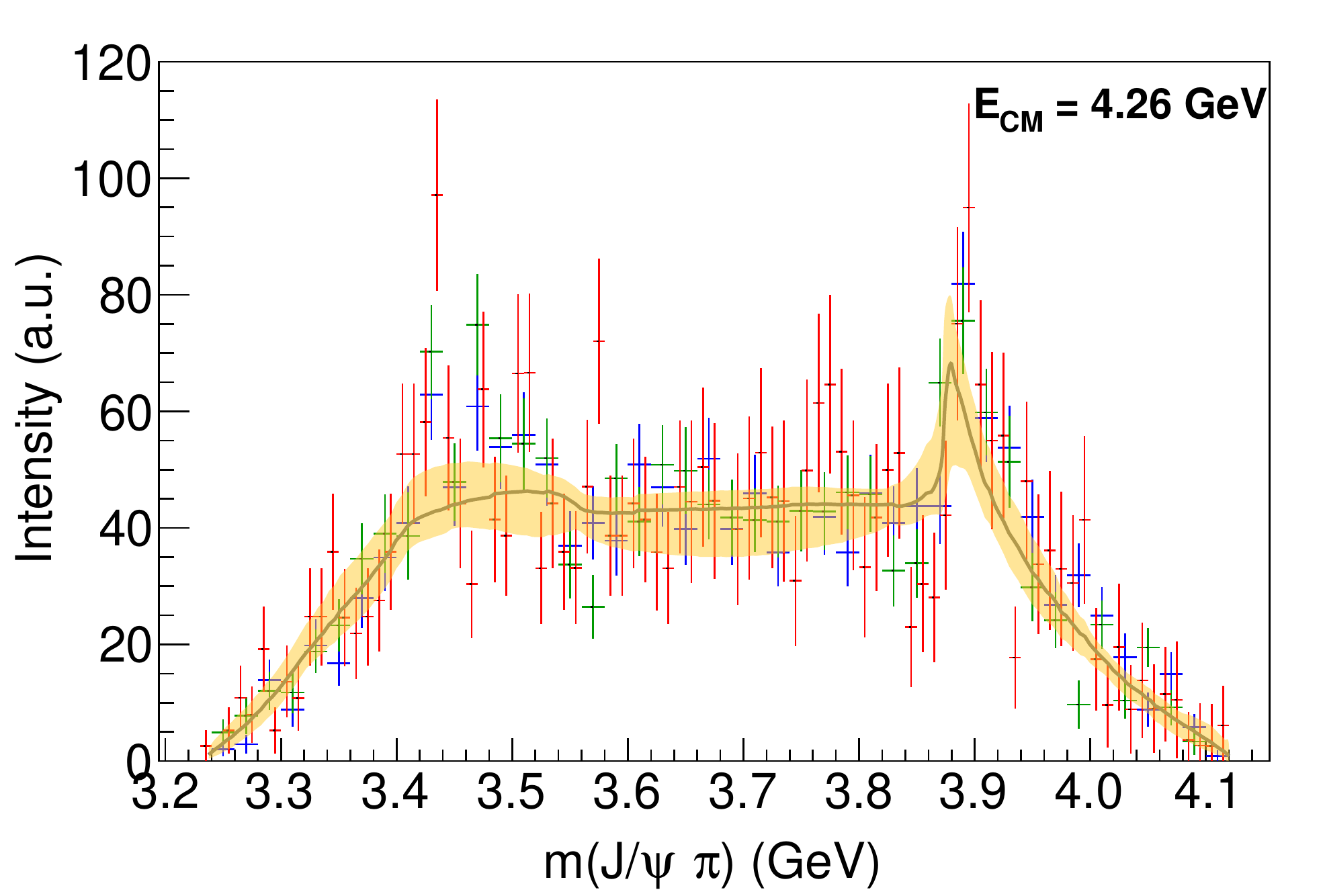}
\caption{\small ~}
\end{subfigure}
 \begin{subfigure}[t]{.32\columnwidth}
\includegraphics[width=\textwidth]{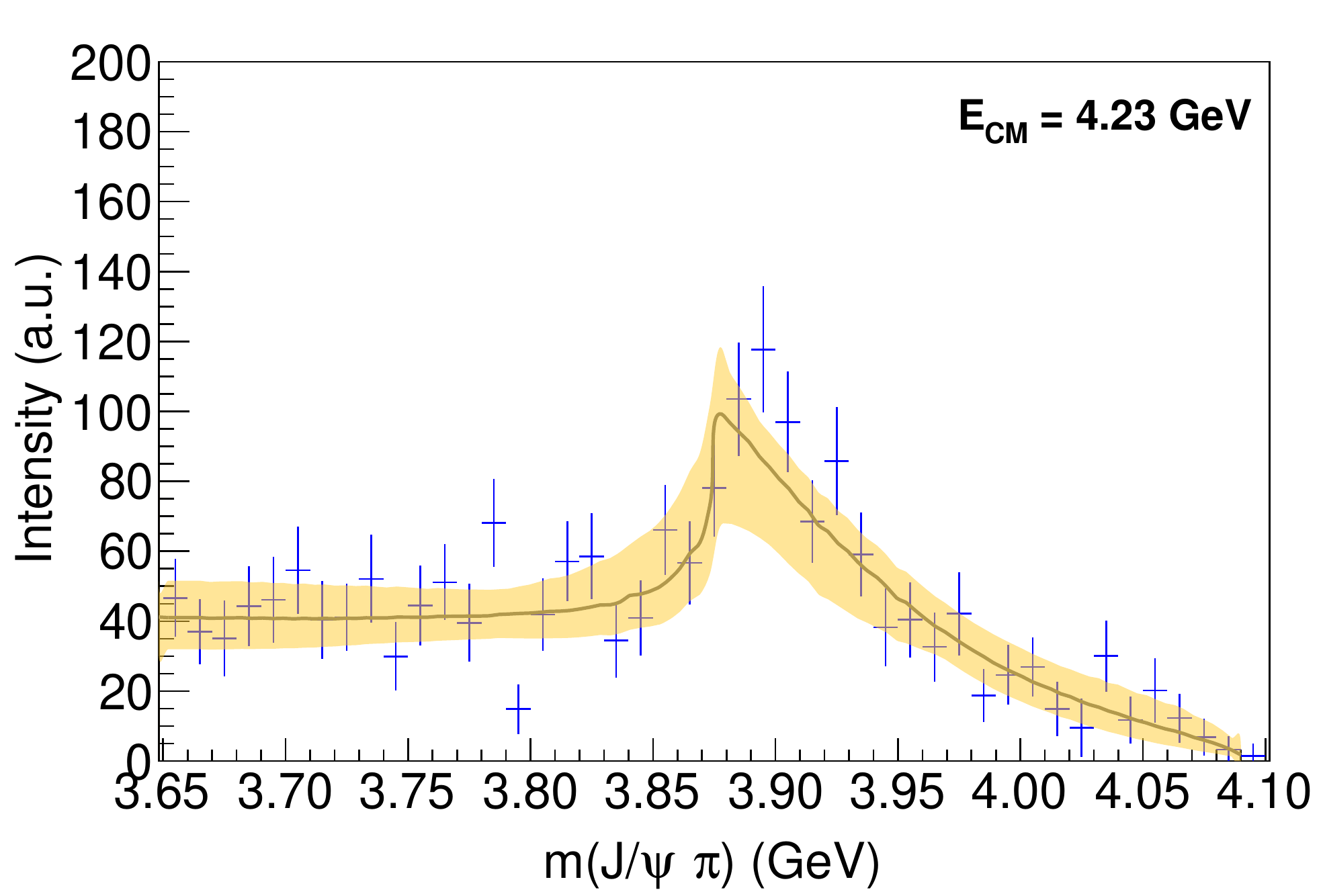}
\caption{\small ~}
\end{subfigure}
 \begin{subfigure}[t]{.32\columnwidth}
\includegraphics[width=\textwidth]{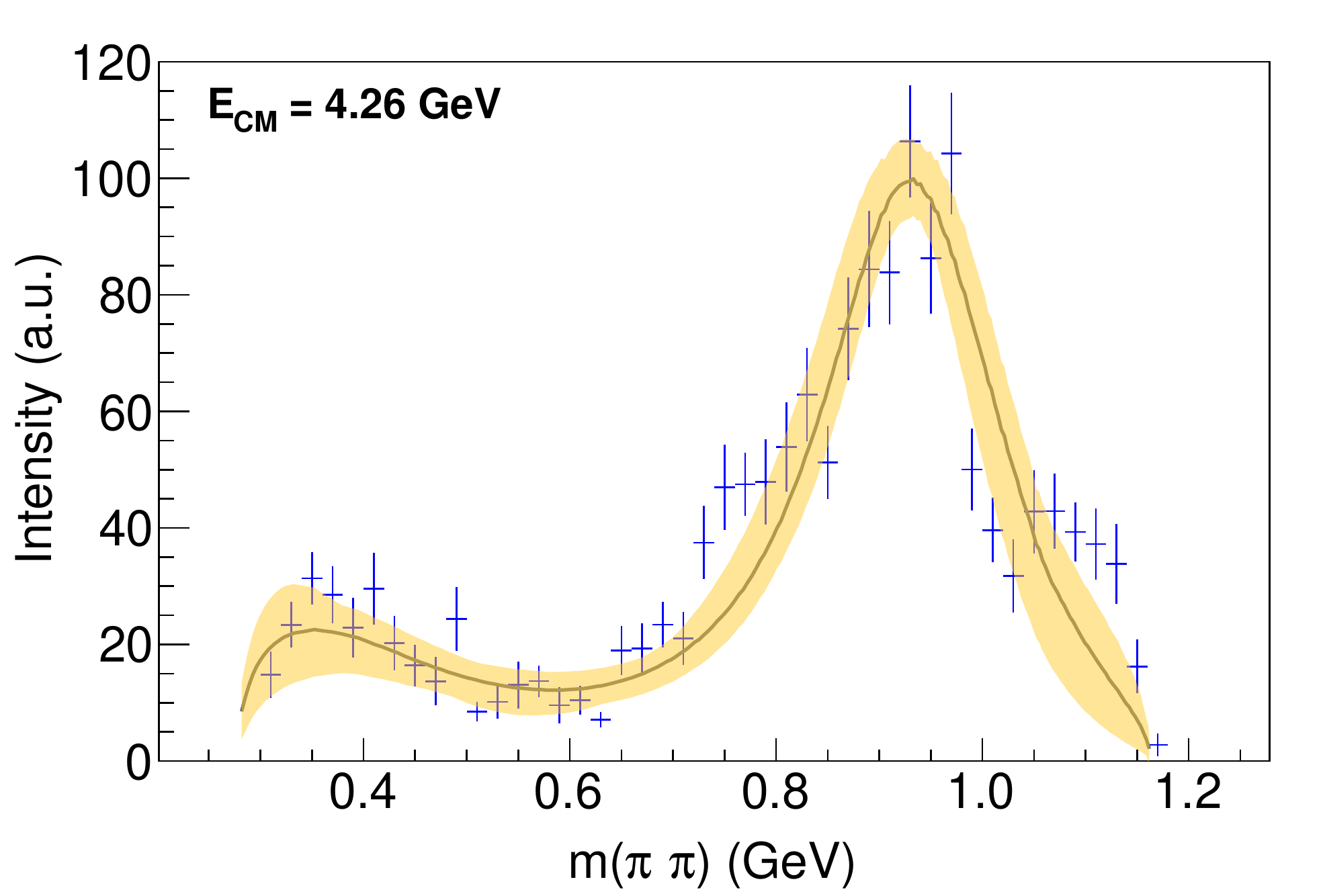}
\caption{\small ~}
\end{subfigure}\\
 \begin{subfigure}[t]{.32\columnwidth}
\includegraphics[width=\textwidth]{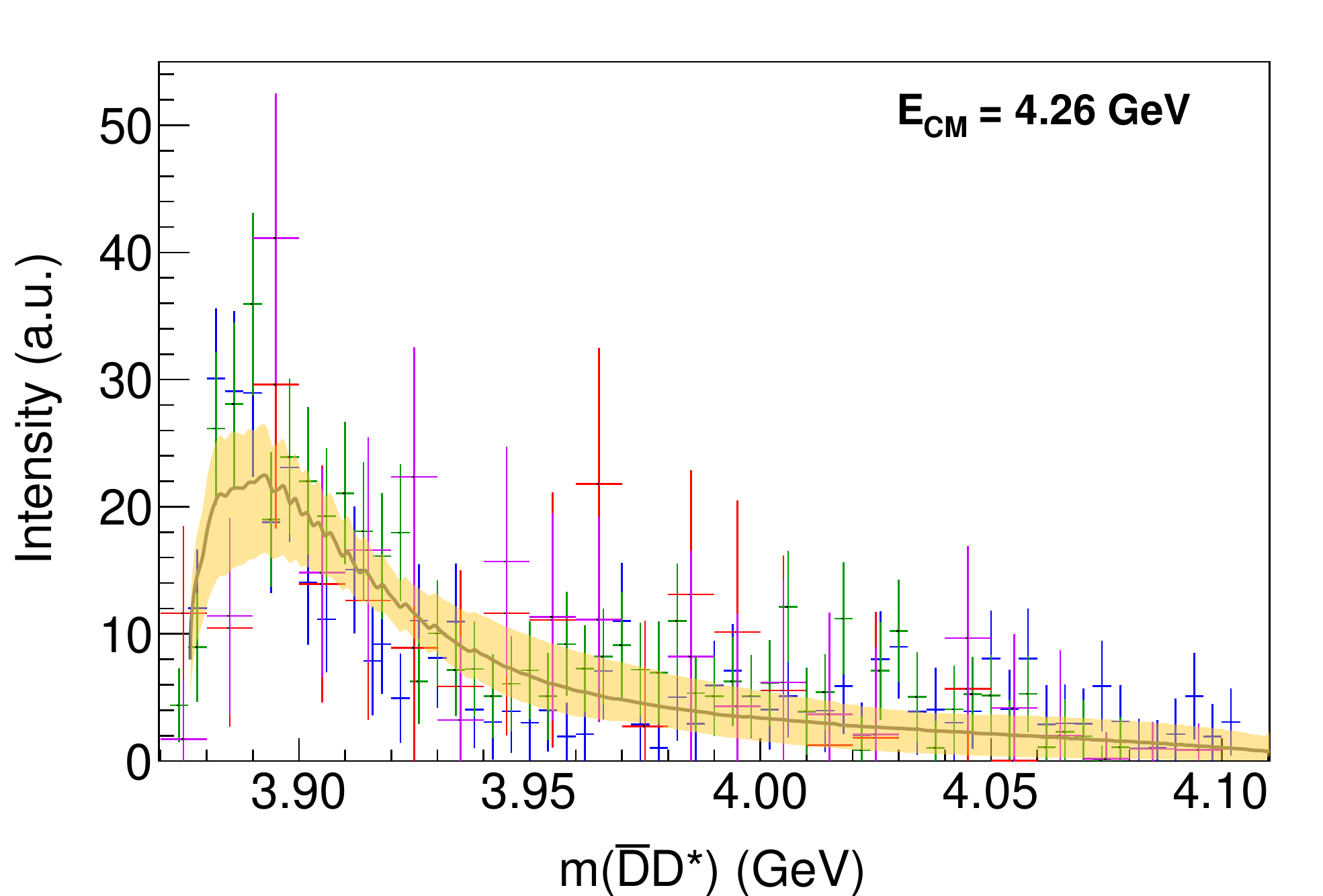}
\caption{\small ~}
\end{subfigure}
 \begin{subfigure}[t]{.32\columnwidth}
\includegraphics[width=\textwidth]{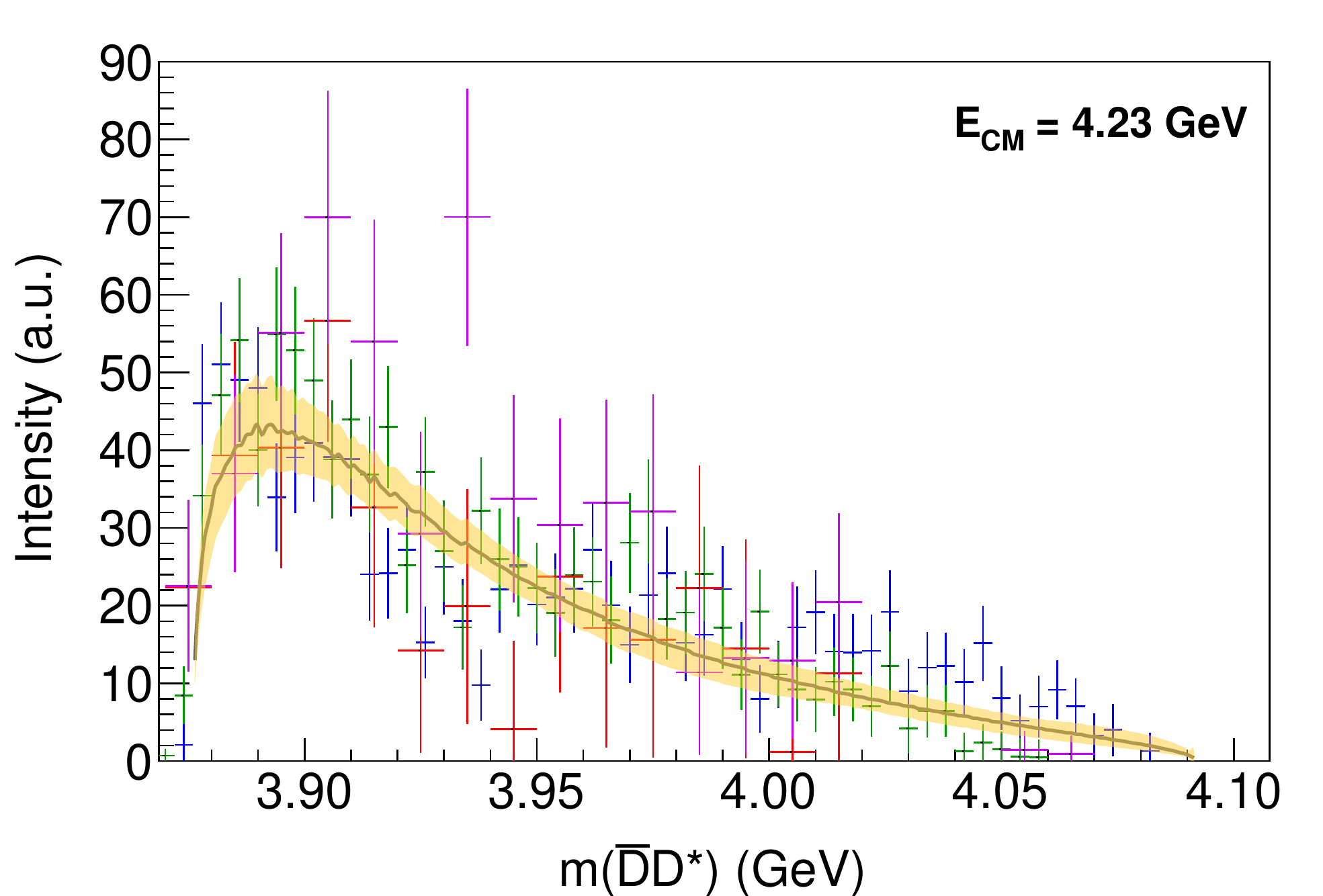}
\caption{\small ~}
\end{subfigure}
 \caption{Result of the fit for the  scenario \scentr (triangle singularity only). The plot legend and the comments on the fit are given in the caption of \figurename{~\ref{fig:scenIII}}. }
 \label{fig:scentr}
 \end{figure}
 
 %%%%%%%%%%%%%%%%%%%%%%%%%%%%%%%%
%	Description of the dataset
%%%%%%%%%%%%%%%%%%%%%%%%%%%%%%%%
\section{Description of the dataset}
All the relevant mass distributions that we discuss are not corrected by acceptance or efficiency. This prevents us from 
 giving  the absolute normalization of our amplitudes, or to quote physical values for the couplings. Most of the experimental analyses include some reducible incoherent background from sidebands or MonteCarlo (MC) simulations, which should be subtracted before comparing with our amplitude prediction. However, in the analyses we consider, this background seems to be rather small ($\sim 15 \%$ of events) and flat~\cite{Ablikim:2013mio,Ablikim:2015swa,Ablikim:2015tbp}. Thus, since we do not give absolute normalizations  of the amplitudes and this background does not affect the shape of the distributions we simply neglect it. The only exception is the neutral $\bar D D^* \pi^0$ channel~\cite{Ablikim:2015gda}, where the mis-reconstructed events are a large fraction of the Dalitz plot and have a nontrivial shape. A curve parametrizing this background, obtained from MC simulations, is shown in ~\cite{Ablikim:2015gda},  but with no associated uncertainties. In our analysis we use the same shape to 
  subtract from the signal and assume Poissonian uncertainties. 

As discussed in the introduction, the $Z_c(3900)$ has been observed in $e^+ e^- \to \jpsi \pi^+ \pi^-$ by \bes at fixed beam energy of $E_\text{CM} = 4260\mev$~\cite{Ablikim:2013mio}. We include in the fit the three projections of the Dalitz plot $m(\jpsi \pi^+)$, $m(\jpsi \pi^-)$, and $m(\pi^+\pi^-)$ quoted in the paper, ignoring their correlations. 
\belle published a similar analysis, but the final state is produced in association with an undetected ISR photon~\cite{Liu:2013dau}. The systematics which affect this observation mode are rather different from those by \bes, and since this dataset is smaller we do not use it. Similarly, we do not consider the low statistics analysis of the CLEO-$c$ data~\cite{Xiao:2013iha}. 
\bes also reported the observation in the neutral channel,  $e^+ e^- \to \jpsi \pi^0 \pi^0$~\cite{Ablikim:2015tbp}. The paper shows only the $m(\jpsi \pi^0)$ projection, at the energies $E_\text{CM} =4230$, $4260$, and $4360\mev$.  The distributions at $4230$ and  $4360\mev$ are shown only  for $m(\jpsi \pi^0) > 3650\mev$. We include in the fit the two datasets at $4230$ and $4260\mev$. To match the charged data, the $4260\mev$ dataset has been rescaled by isospin, binning, and efficiency (with the values quoted in~\cite{Ablikim:2013mio,Ablikim:2015tbp}).

For the open charm channel, we consider the double-tag analysis by \bes~\cite{Ablikim:2015swa} of $e^+e^- \to \bar D^0 D^{*+} \pi^-$ and $e^+e^- \to \bar D^{*0} D^{+} \pi^-$. The paper quotes the mass projection $m(\bar D D^*)$ only, at the energies $E_\text{CM} = 4230$ and $4260\mev$. We include in the fit all four datasets. The previous \bes single-tag analysis~\cite{Ablikim:2013xfr} is somehow statistically independent from the latter, but the data are affected by larger incoherent backgrounds from  mis-reconstructed $D^{(*)}$ mesons, and we do not include them in the fit. Instead, we consider the four $m(\bar D D^*)$ distributions of the neutral channel $e^+e^- \to \bar D^0 D^{*0} \pi^0$ and $e^+e^- \to D^{-} D^{*+} \pi^0$, at  energies $E_\text{CM} = 4230$ and  $4260\mev$. We subtract the incoherent background from these data, then we rescale to match the number of events in the charged channel, to take into account the unquoted different efficiencies.

Our full dataset has $566$ experimental points. We work in the isospin symmetric limit, so we use $m_\pi = (2 m_{\pi^+} + m_{\pi^0})/3$, and $m_{D^{(*)}} = \left(m_{D^{(*)0}} + m_{D^{(*)+}}\right)/2$. Four points happen to be below the iso-symmetric $\bar D D^*$ threshold, and are removed from the fit.
The values of masses and widths of the final state mesons, and of the intermediate $D_1(2420)$ and $D_0(2400)$ are taken from the Particle Data Group (PDG)~\cite{pdg}. Since we are not directly interested in the $m(\pi\pi)$ distribution, we parametrize the $\pi\pi$ resonances with ``effective'' $f_0(980)$ and $\sigma$, whose masses and widths are $M_{f_0}=920\mev$, $\Gamma_{f_0}=223\mev$, and $M_\sigma=112\mev$, $\Gamma_\sigma=906\mev$, respectively, as determined from a preliminary fit to the $\pi\pi$ distribution only. The data points at $m(\pi\pi) > 1\gev$ are not well described by this choice, and since they do not affect the $\jpsi \pi$ and $\bar DD^*$ distribution we are interested in, we remove these points from the fit.
This parametrization breaks unitarity in the $t$-channel, and does not include either the $K\bar K$ channel (important to have a good description of the $f_0$), or the Adler zero (which generates the $\sigma$ pole). Nevertheless is good enough to describe the the projection in the $s$-channel. We remark that, because of these approximations, the Breit-Wigner parameters we quote for the $\sigma$ and $f_0$ can by no means be compared with the right ones extracted with more refined techniques. 
Higher statistics will require a thorough parametrization of the $\pi\pi$ scattering~\cite{GarciaMartin:2011cn,Bydzovsky:2016vdx}.

We consider the datasets at the different center-of-mass energies as independent samples, that is couplings at different $E_\text{CM}$ are independent fit parameters. Thus our model has 15 fit parameters:  for each one of the two center-of-mass energies, the amplitudes in Eqs.~\eqref{eq:ampl1} and~\eqref{eq:ampl2} have one coupling for each one of the 4 exchanged resonances, and the two short-range coefficients (subtraction constants). Both center-of-mass energies share the same $K$ matrix, which is parametrized with three constants $K_{11}$, $K_{12}$, and $K_{22}$ in the scenarios \scenIV and \scentr, and by two couplings and a mass ($g_1$, $g_2$, $M$) in the scenarios \scenIII and \scenIIItr. Considering this, and the number of points removed from the dataset, the four fits have 532 degrees of freedom.

\begin{table}[b]
\centering
\begin{tabular}{l|ccc}
Scenario & $\chi^2$ & DOF & $\chi^2/\text{DOF}$ \\ \hline
\scenIII & $644$ & $532$ & $1.21$ \\
\scenIIItr & $642$ & $532$ & $1.21$ \\
\scenIV & $666$ & $532$ & $1.25$ \\
\scentr & $695$ & $532$ & $1.31$ \\ 
\end{tabular}
\caption{Best fit $\chi^2$ for all the different scenarios examined.} 
\label{tab:chisq}
\end{table}
\begin{figure}[p]
\centering
 \begin{subfigure}[t]{.30\columnwidth}
\includegraphics[width=\textwidth]{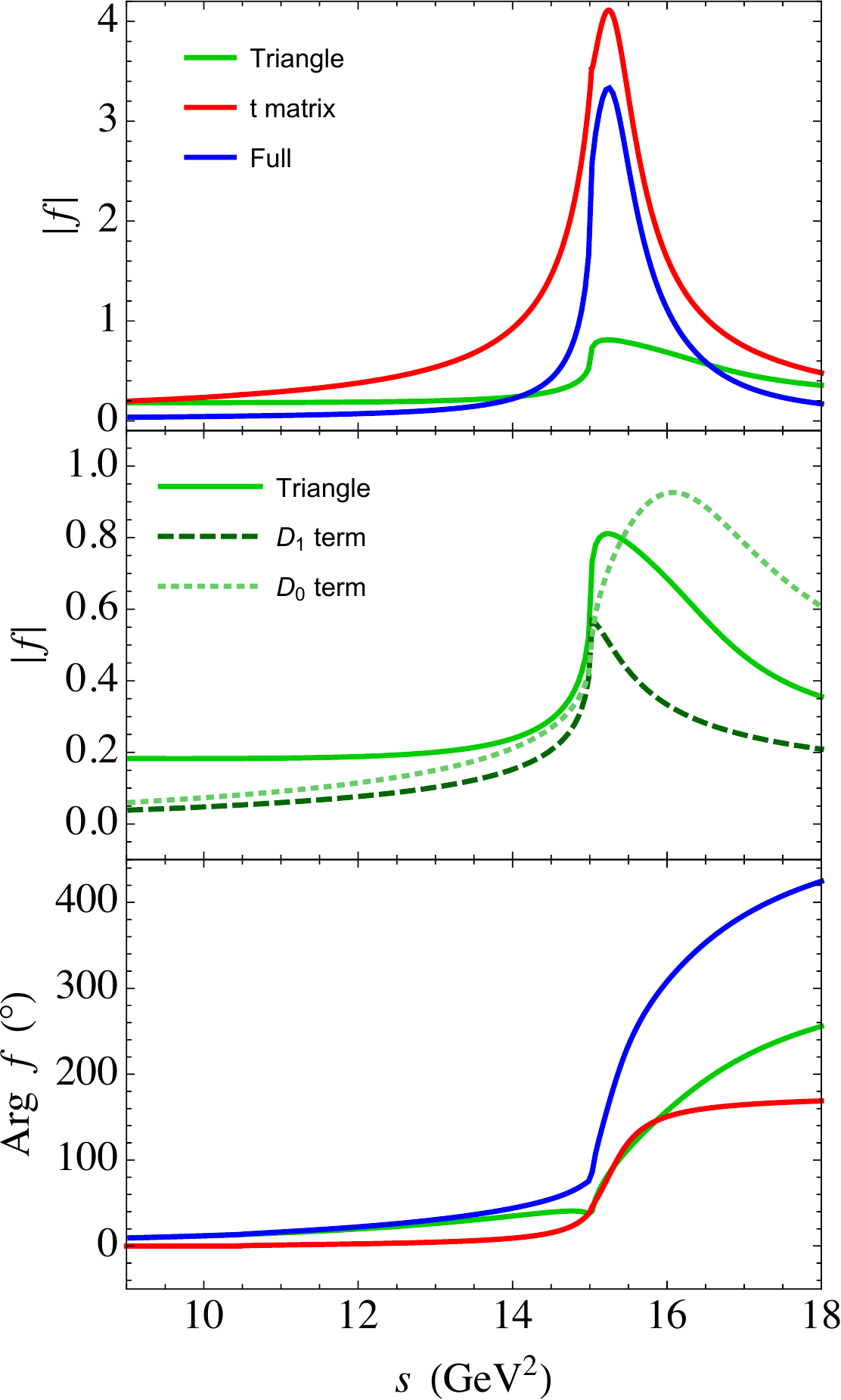}
\caption{\small \scenIIItr}
\end{subfigure}
 \begin{subfigure}[t]{.30\columnwidth}
\includegraphics[width=\textwidth]{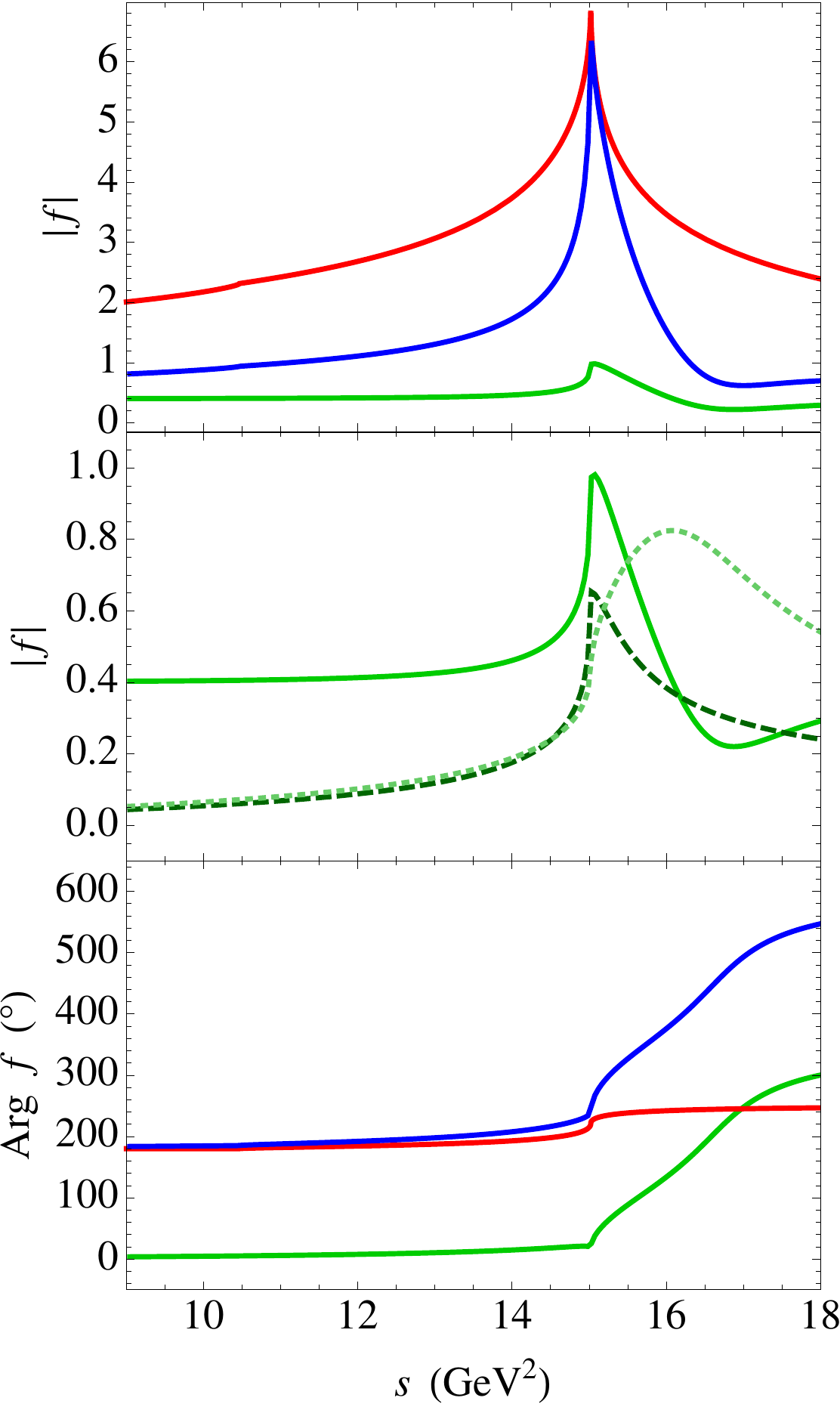}
\caption{\small \scenIV}
\end{subfigure}
 \begin{subfigure}[t]{.30\columnwidth}
\includegraphics[width=\textwidth]{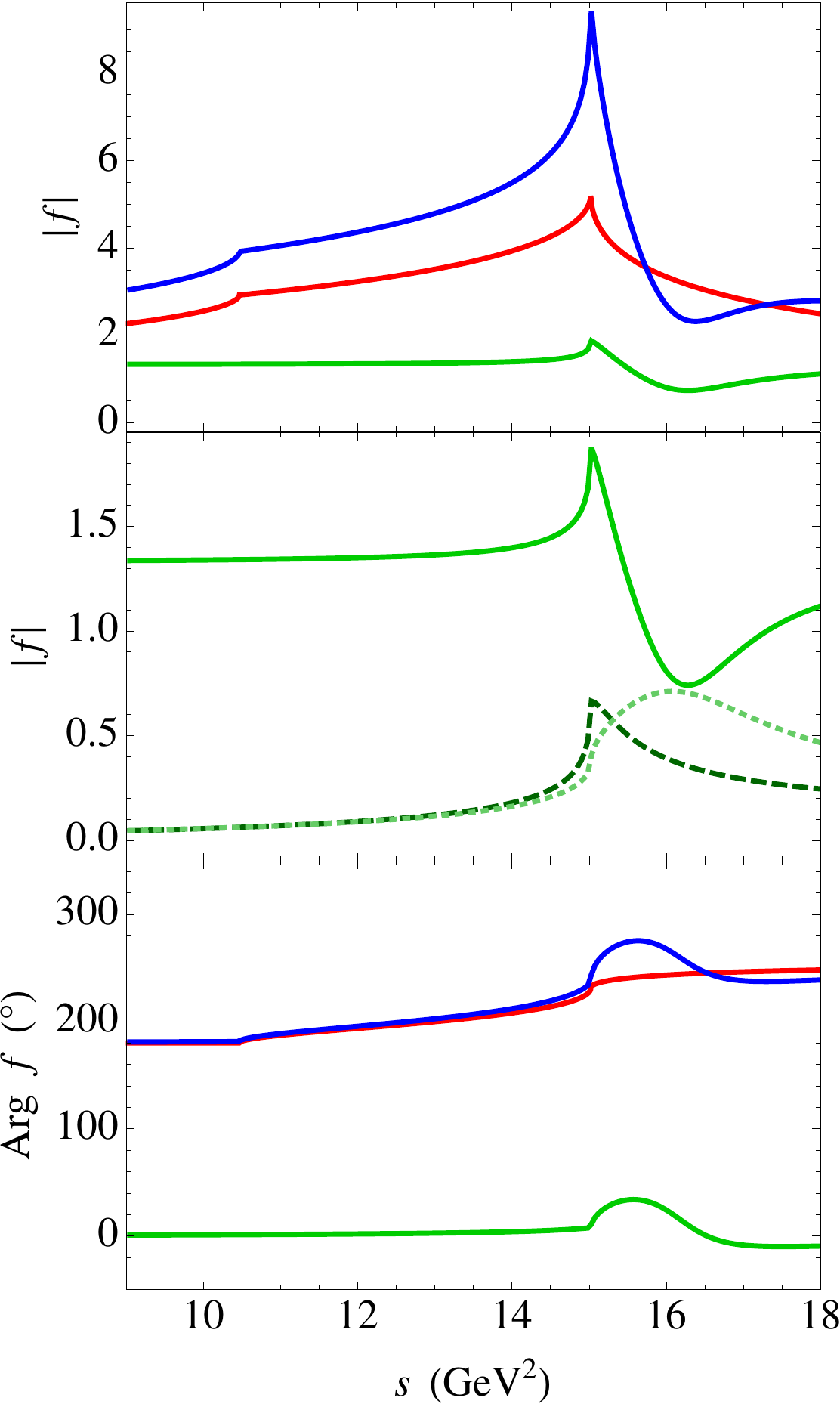}
\caption{\small \scentr}
\end{subfigure}
\caption{Interplay of scattering amplitude poles and triangle singularity to reconstruct the peak. We focus on the $\jpsi \pi$ channel, at $E_{CM} = 4.26$~GeV. The red curve is the $t_{12}$ scattering amplitude, the green curve is the $c_1 + H(s,D_1) +  + H(s,D_0)$ term in Eq.~\eqref{eq:ampl2}, and the blue curve is the product of the two. The upper plots show the magnitudes of these terms, the lower plots the phases. The middle row shows the contributions to the unitarized term due to the $D_1$ (dashed) and the $D_0$ (dotted). Only for $D_1$ the singularity is close enough to the physical region to generate a large peak. (a) The pole on the III sheet generates a narrow Breit-Wigner-like peak. The contribution of the triangle is not particularly relevant. (b) The sharp cusp in the scattering amplitude is due to the IV sheet pole close by; the triangle contributes to make the peak sharper. (c) The scattering amplitude has a small cusp due to the threshold factor, and the triangle is needed to make it sharp enough to fit the data. }
\label{fig:interplay}

\vspace{.5cm}
 \begin{subfigure}[t]{.30\columnwidth}
\includegraphics[width=\textwidth]{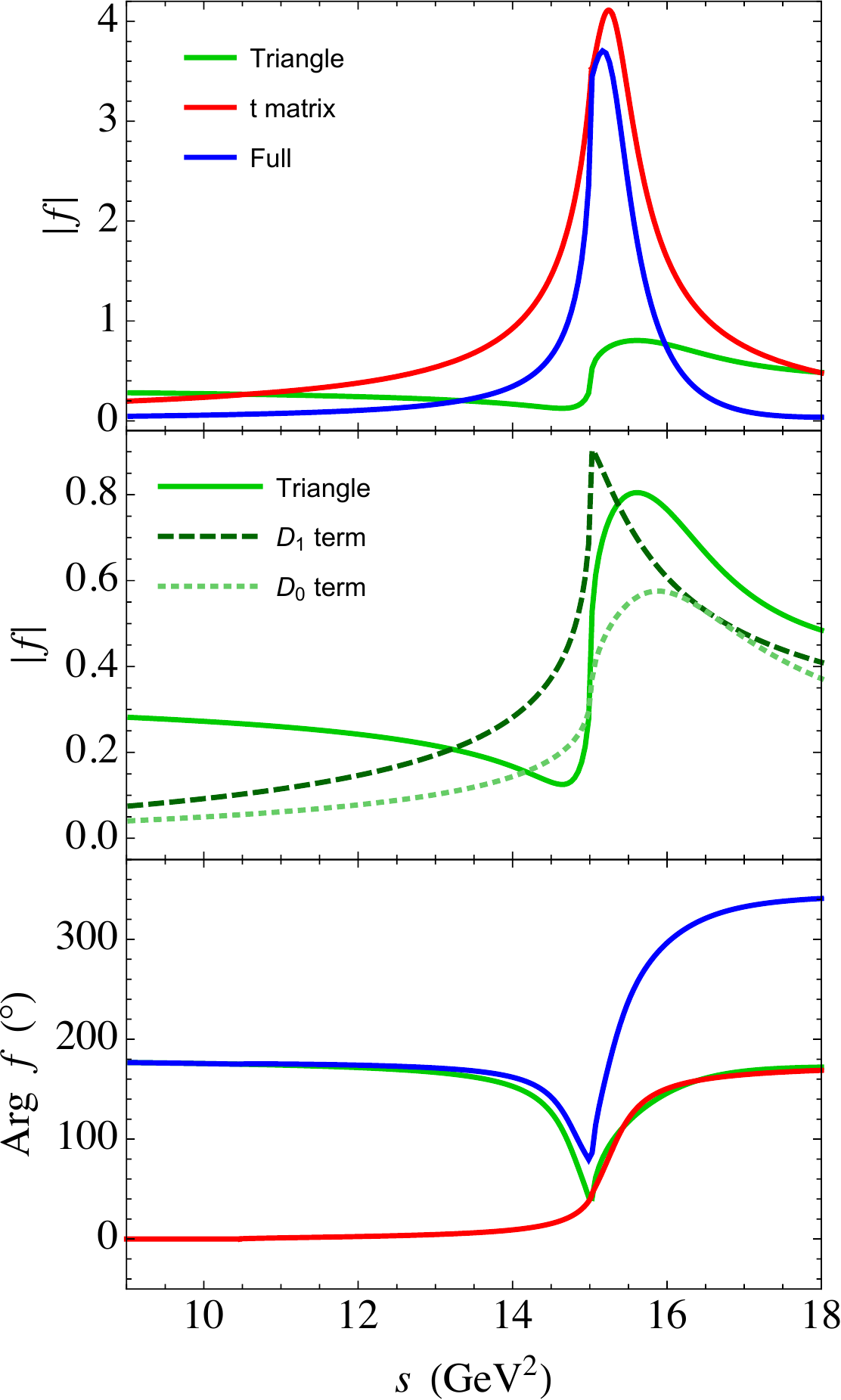}
\caption{\small \scenIIItr}
\end{subfigure}
 \begin{subfigure}[t]{.30\columnwidth}
\includegraphics[width=\textwidth]{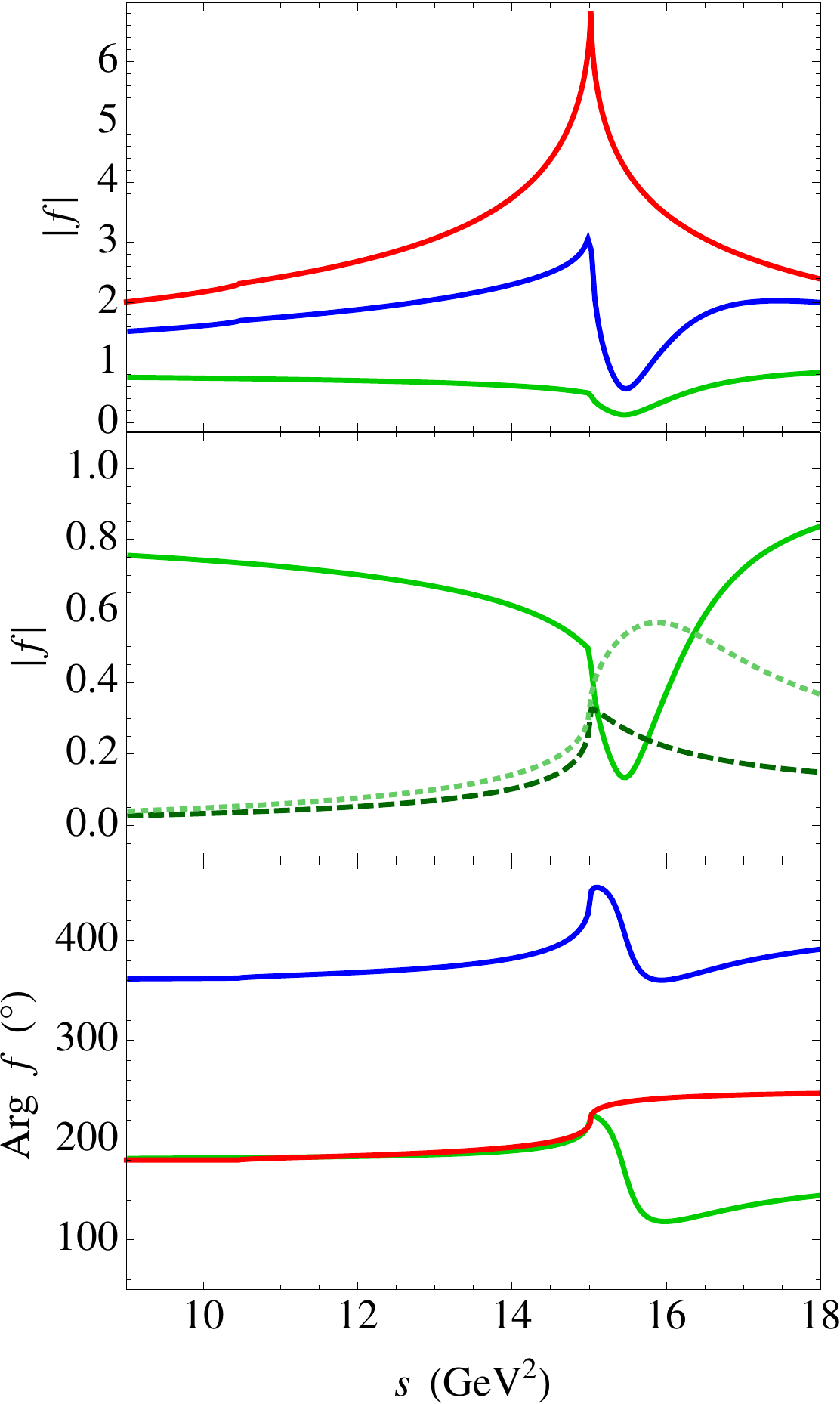}
\caption{\small \scenIV}
\end{subfigure}
 \begin{subfigure}[t]{.30\columnwidth}
\includegraphics[width=\textwidth]{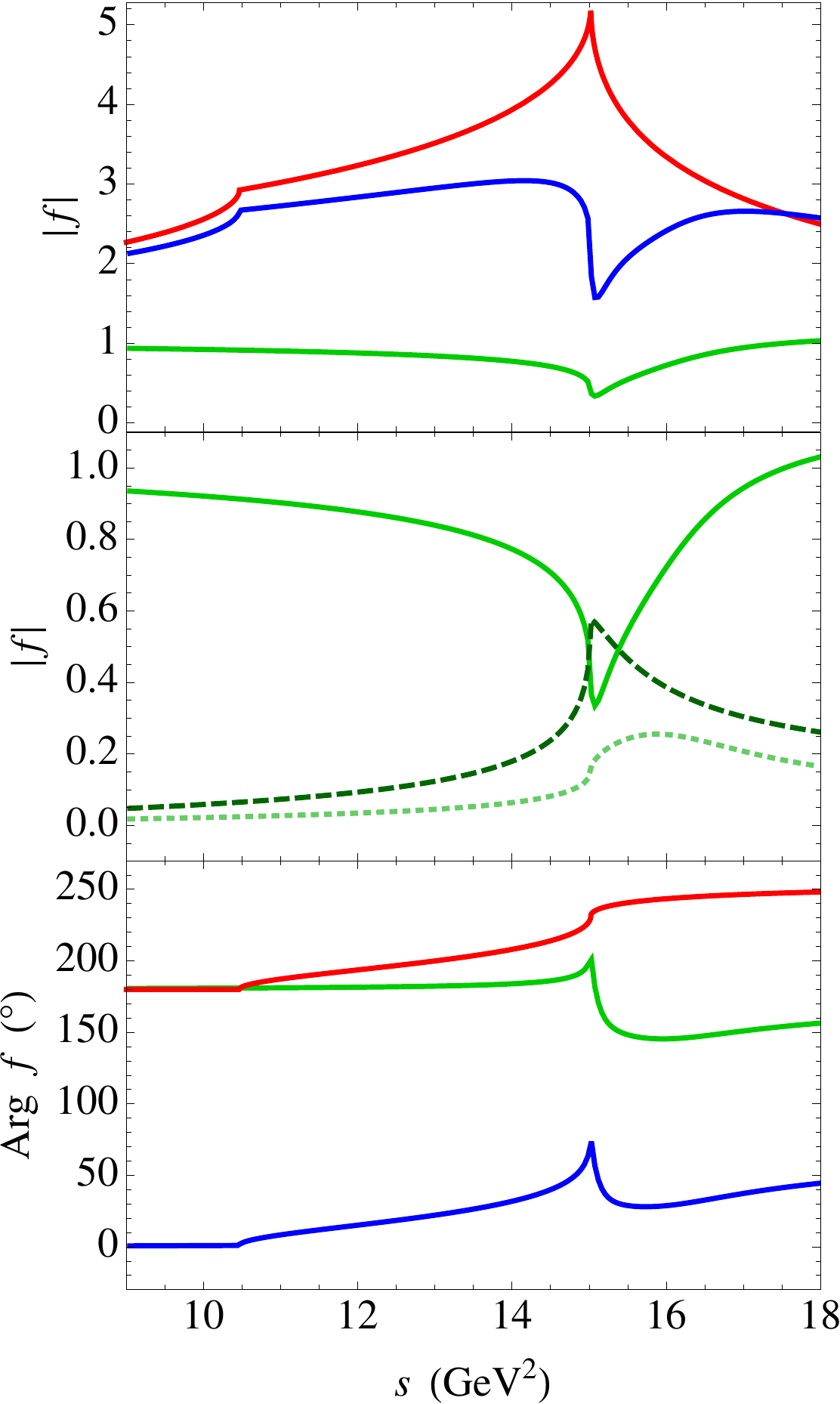}
\caption{\small \scentr}
\end{subfigure}
\caption{Same as \figurename{~\ref{fig:interplay}}, but for $E_{CM} = 4.23$~GeV. }
\label{fig:interplay2}
\end{figure}
\section{Fit results}
 \label{sec:fits}
We perform a minimum $\chi^2$ fit using \textsc{minuit}~\cite{minuit}. In Figures~\ref{fig:scenIII}, \ref{fig:scenIIItr}, \ref{fig:scenIV}, and \ref{fig:scentr} we show the results of the fits for the four scenarios. The starting values of the fit parameters have been set by looking for the best $\chi^2$ of $O(10^4)$ preliminary fits with randomly chosen initial parameters. 
The mean value and uncertainty of the fitted curve
have been computed using the bootstrap technique~\cite{recipes,Blin:2016dlf,Landay:2016cjw}, which allows us to take into
account correlations among fit parameters and to properly
propagate the uncertainties not only to all the observables 
but also to any quantity that can be extracted from the amplitude (\eg the pole positions).
Specifically, for each one of the four models, we generate 2000 datasets
by randomly sampling the experimental points according to Gaussian distributions. 
For each pseudo-dataset, we perform an independent minimum $\chi^2$ fit, using as initial conditions the ones of the original fit. We can thus select the best
68\% fits ($1\sigma$ confidence level). The $\chi^2$s of the best fits are reported in \tablename{\ref{tab:chisq}}.
 
All the models have a $\chi^2/\text{DOF} \sim 1.3$, and give a rather good description of the dataset, as can be seen 
 from Figures~\ref{fig:scenIII}, \ref{fig:scenIIItr}, \ref{fig:scenIV}, and \ref{fig:scentr}. The peak at $\sqrt{s} \simeq 3.4$~GeV is due to the reflection of the structure at right, and cannot be reproduced properly if spins are neglected. To show separately the contribution of the  triangle singularity and of the pole in the scattering matrix, in \figurename{~\ref{fig:interplay} and~\ref{fig:interplay2}} we show the magnitude and the phase of the $t_{12}$ amplitude, of the unitarized term $c_1 + H(s, D_1) + H(s, D_0)$, and the product of the two. We also plot separately the contributions of the $D_0$ and $D_1$ exchanges. We show that the latter only is in the kinematical regime to generate a sharp peak close to the physical region.

 \begin{figure}[t]
\centering
\includegraphics[width=.32\textwidth]{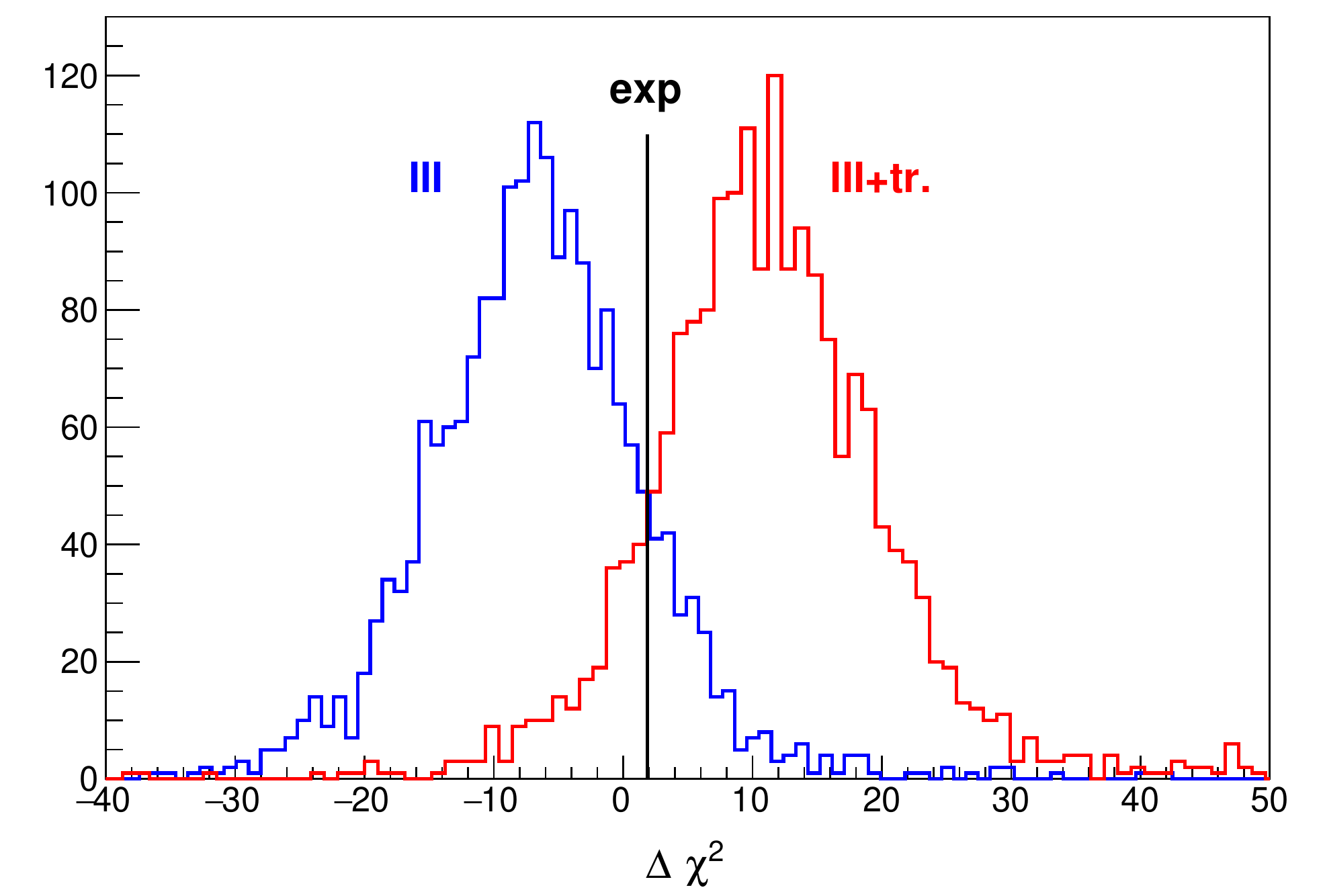}
\includegraphics[width=.32\textwidth]{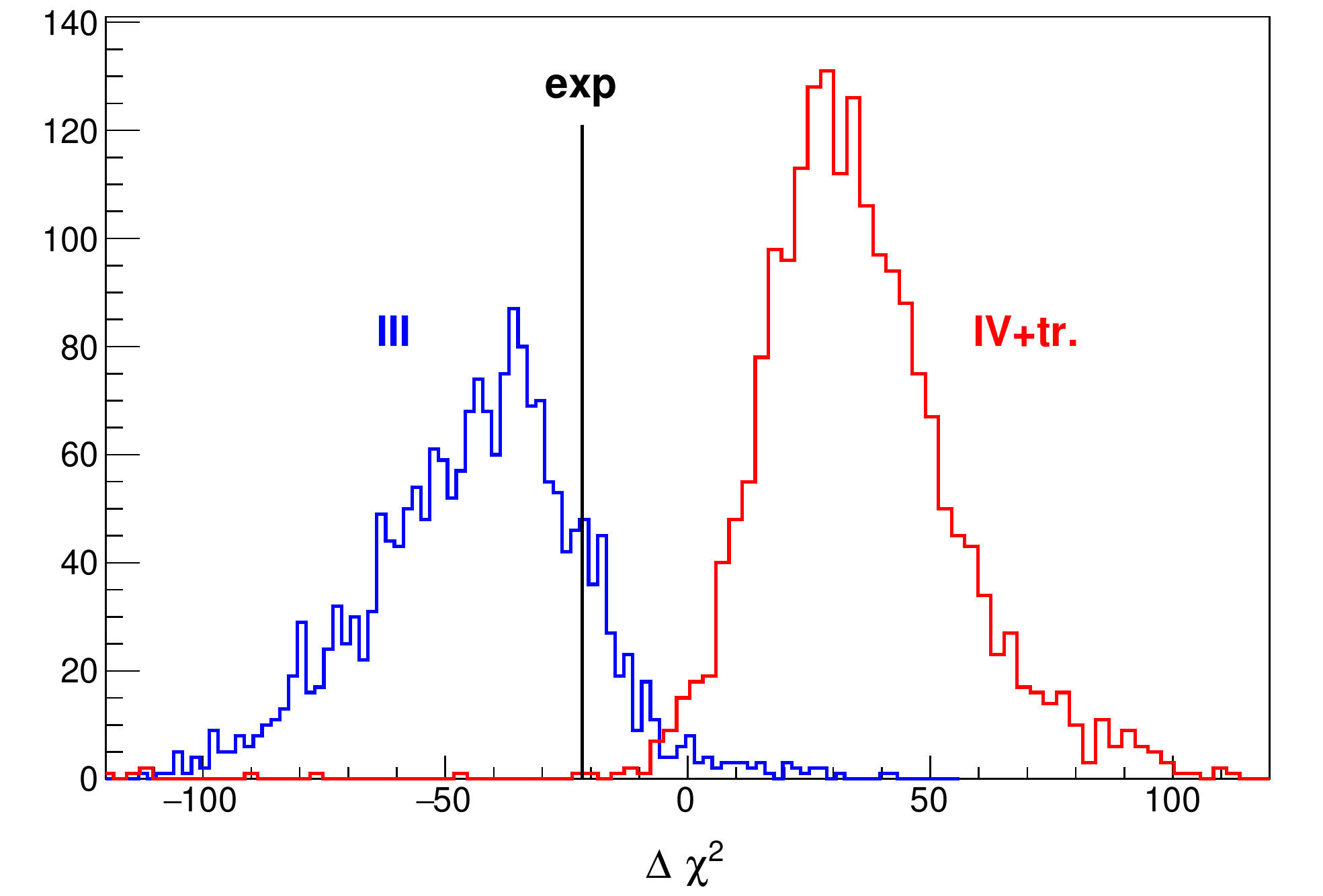}
\includegraphics[width=.32\textwidth]{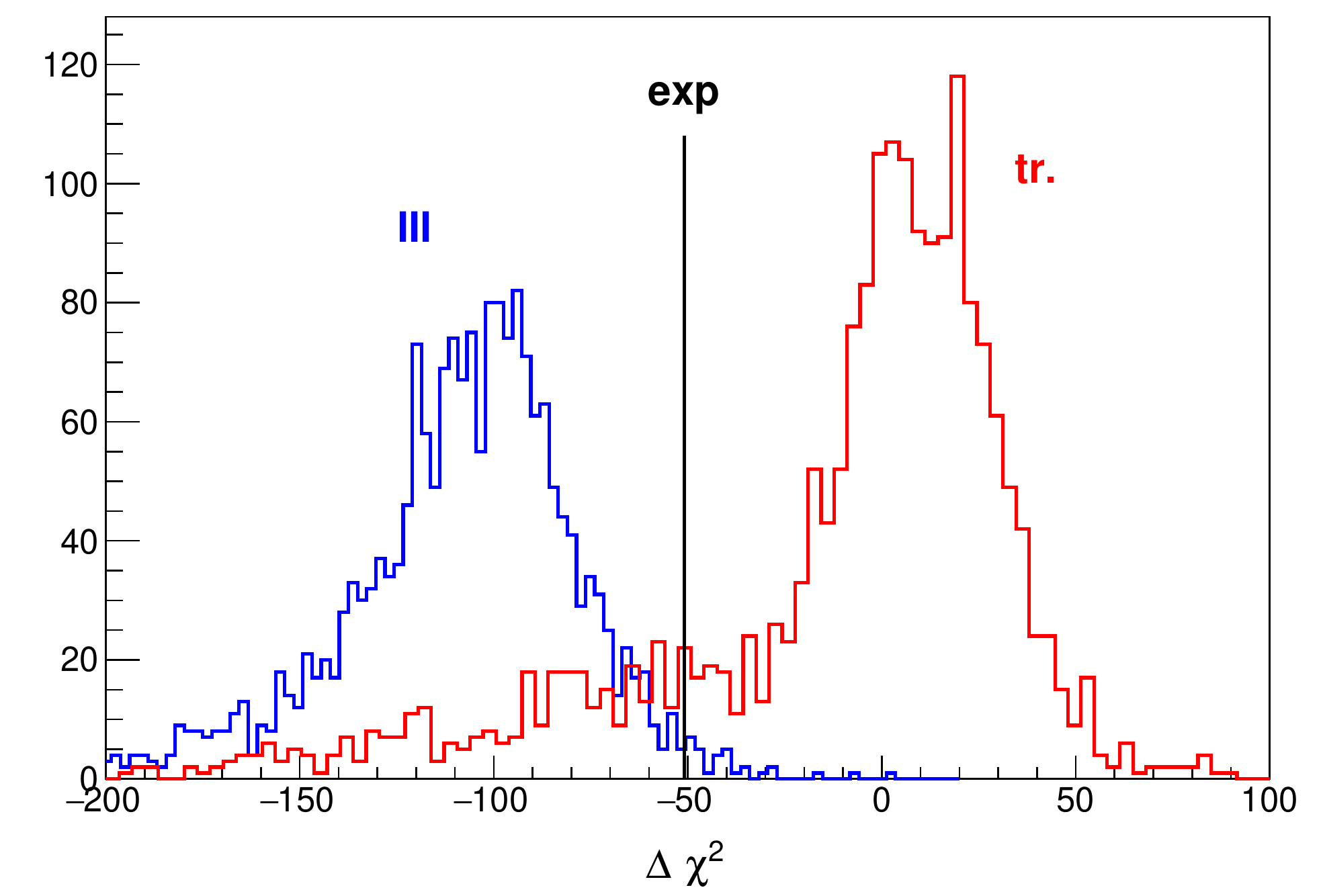}\\
\includegraphics[width=.32\textwidth]{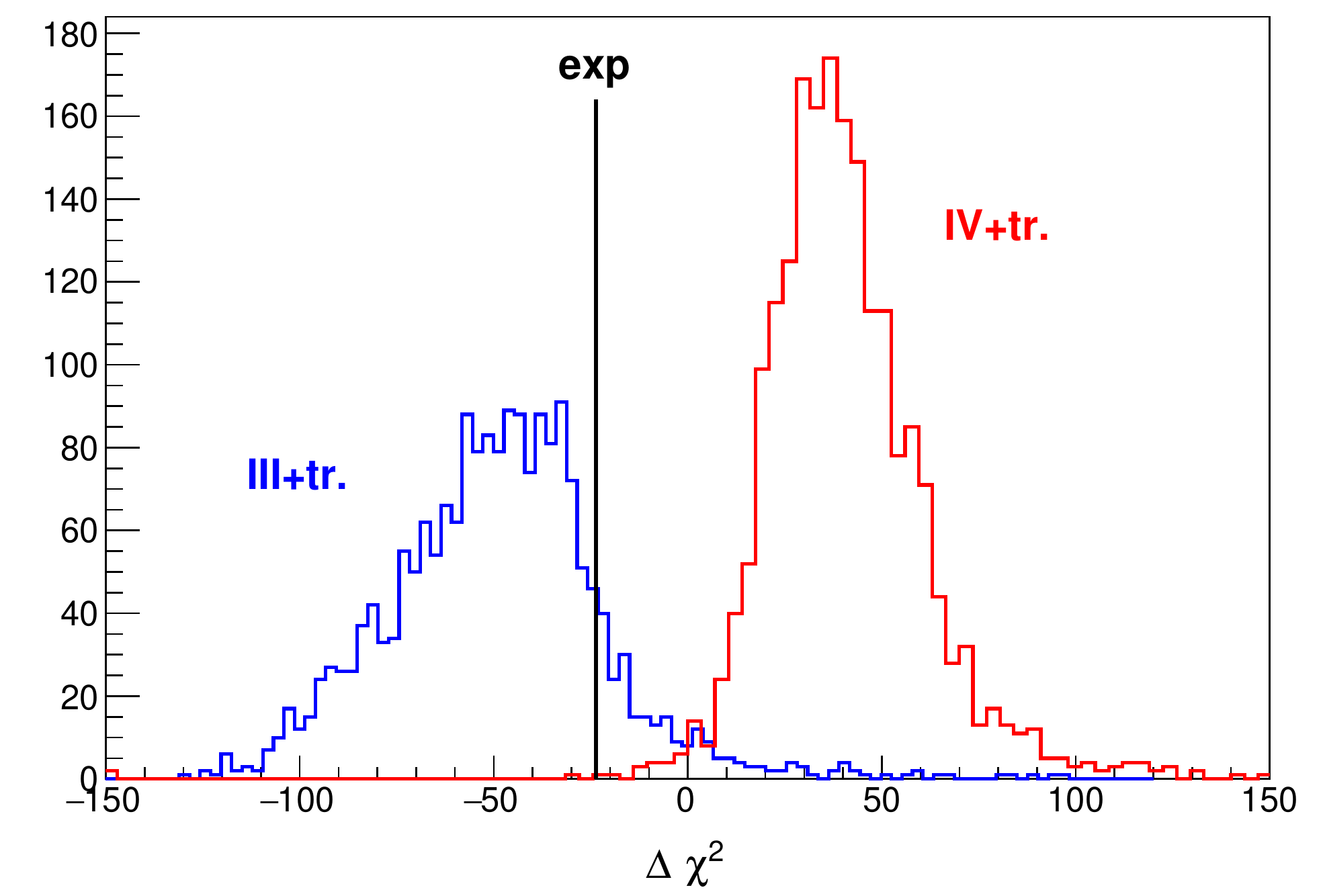}
\includegraphics[width=.32\textwidth]{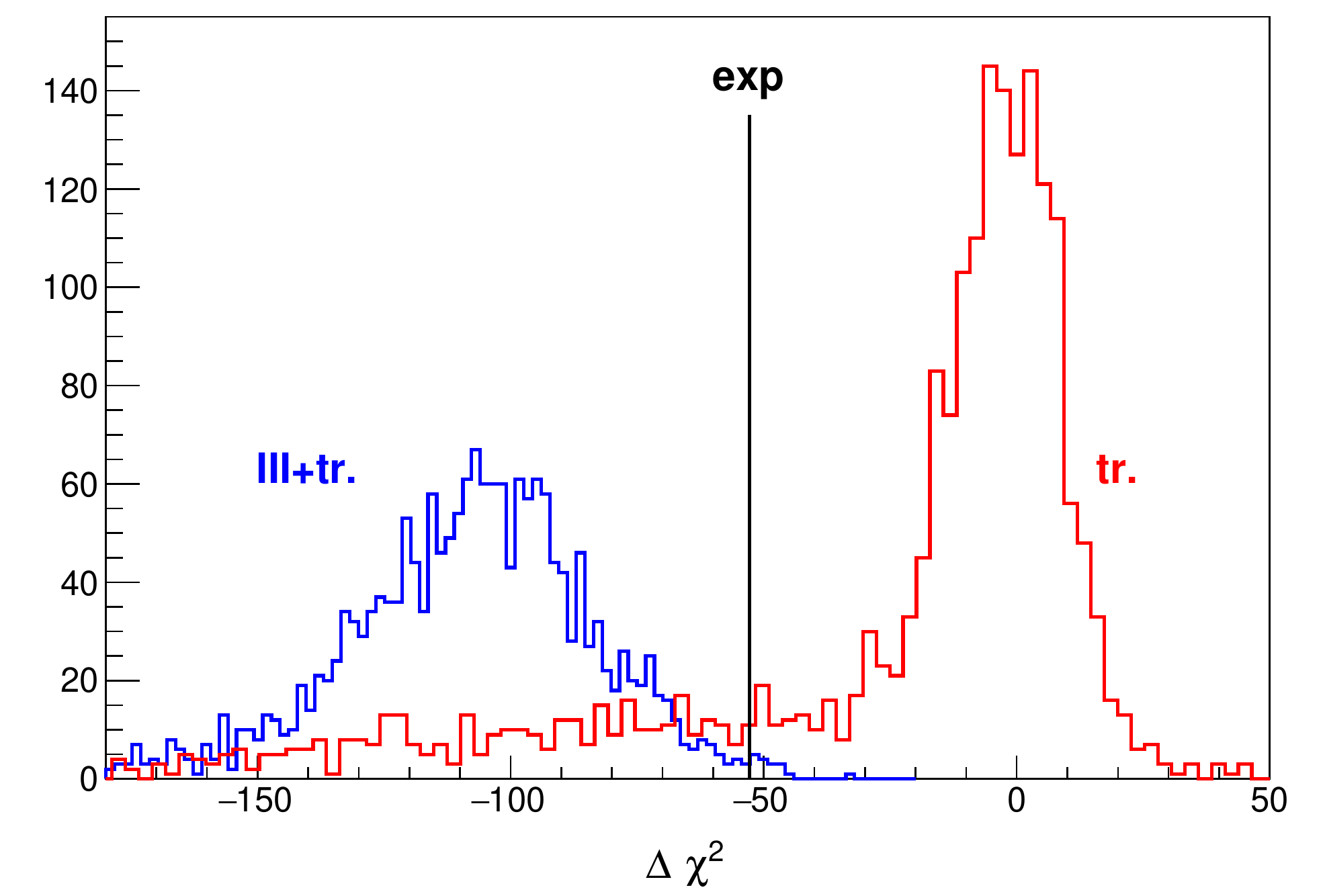}
\includegraphics[width=.32\textwidth]{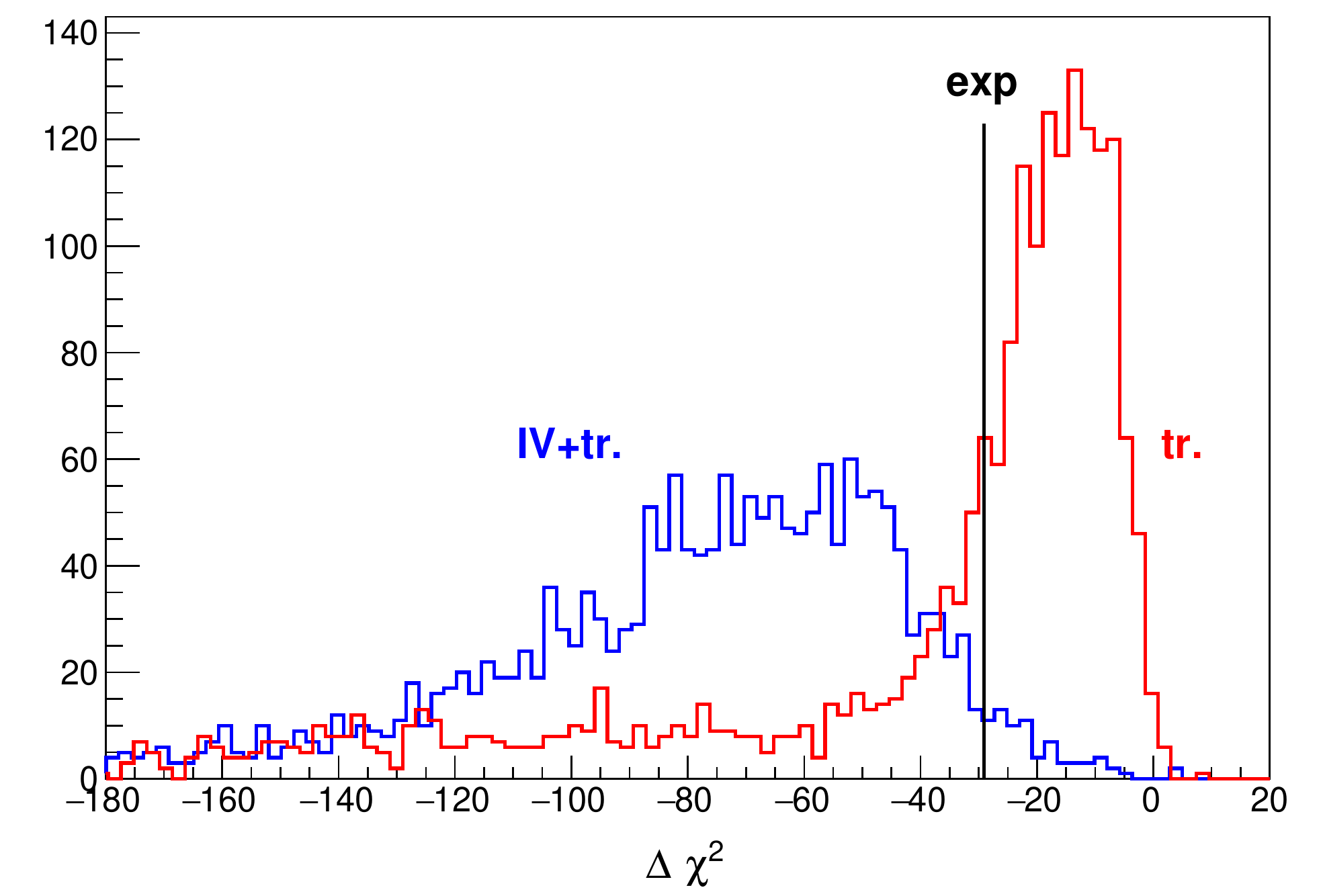}
\caption{Histograms of $\Delta\chi^2$, as explained in the text, that give the significance of each scenario with respect to another one. The significances involving the \scentr option are affected by the penalty in the $\chi^2$ and have to be considered as mere indications.}
\label{fig:delta}
\end{figure}
 The best $\chi^2$ is obtained with model  \scenIIItr, but the difference with the other models does not seem significant. To properly compare the quality of 
  one model fit vs another  we use the $\Delta \chi^2$ estimator on a number of MC generated datasets~\cite{Demortier:2007zz,Faccini:2012zv}. More specifically,  to give the significance, say of model $A$ with respect to model $B$, we generate 2000 pseudodata samples according to either one of the two models, with the fit parameters obtained from the best fit discussed above. Each data point is generated according to a Poissonian distribution, whose mean value is given by the value of the theoretical model at the center of the bin. Each generated dataset is fitted again with both models, and we fill a histogram with the $\Delta \chi^2 = \chi^2 (A) - \chi^2(B)$ estimator~\footnote{which is equivalent to a likelihood-ratio test, if one assumes Gaussian errors.}. We can thus compare the distribution of $\Delta \chi^2$ of the datasets generated according to $A$ (which is expected to peak at negative values of $\Delta \chi^2$), with the distribution of $\Delta \chi^2$ of the datasets generated according to $B$ (which is expected to peak at positive values of $\Delta \chi^2$). These distributions are used to calculate the fraction of samples in which $\Delta \chi^2$ has a value larger (for $A$) or smaller (for $B$) than the one obtained from data, which can be translated into Gaussian significance. The $\Delta\chi^2$ histograms are shown in \figurename{~\ref{fig:delta}}, and the significances are listed in \tablename{\ref{tab:delta}}. We can appreciate the peculiar behaviour of the $\Delta \chi^2$ distributions for the \scentr model, which peaks at $\Delta\chi^2 \simeq 0$ and exhibits a long tail towards negative values. This is due to the penalty introduced in the $\chi^2$ to push the pole far into the complex plane, thus affecting the pure statistical meaning of the $\chi^2$. The significances relative to the \scentr scenario have thus to be considered as mere indications (that is why in \tablename{\ref{tab:delta}} we report them under quotation marks).  Anyway, we note that all the significances are never greater than $3\sigma$. These are going to be even more diluted if we were to consider the systematic uncertainties. We conclude that present statistics prevents us from drawing any strong statements, but the robustness of the tools we have discussed here will allow us to distinguish the different phenomenological models, when new data will be available.
\begin{table}[b]
\centering
\begin{tabular}{l|ccccc}
Scenario &  \scenIIItr & \scenIV & \scentr \\ \hline
\scenIII &  $1.5\sigma$  ($1.5\sigma$) & $1.5\sigma$ ($2.7\sigma$) & ``$2.4\sigma$'' (``$1.4\sigma$'')\\
\scenIIItr &  $-$ & $1.5\sigma$ ($3.1\sigma$) & ``$2.6\sigma$'' (``$1.3\sigma$'') \\
\scenIV & $-$ & $-$ & ``$2.1\sigma$'' (``$0.9\sigma$'') \\
\end{tabular}
\caption{Significance of each model versus another. The number in the cell $AB$ indicates the probability for the $\Delta\chi^2$ generated according to $A$ ($B$) to be greater (smaller) than the $\Delta\chi^2$ obtained from the fit to the real data. The significances relative to the \scentr option are affected by the penalty in the $\chi^2$ and should be considered as mere indications. } 
\label{tab:delta}
\end{table}

\section{Pole searches}
  \begin{figure}[t]
 \centering
 \begin{subfigure}[b]{.32\textwidth}
\includegraphics[width=\textwidth]{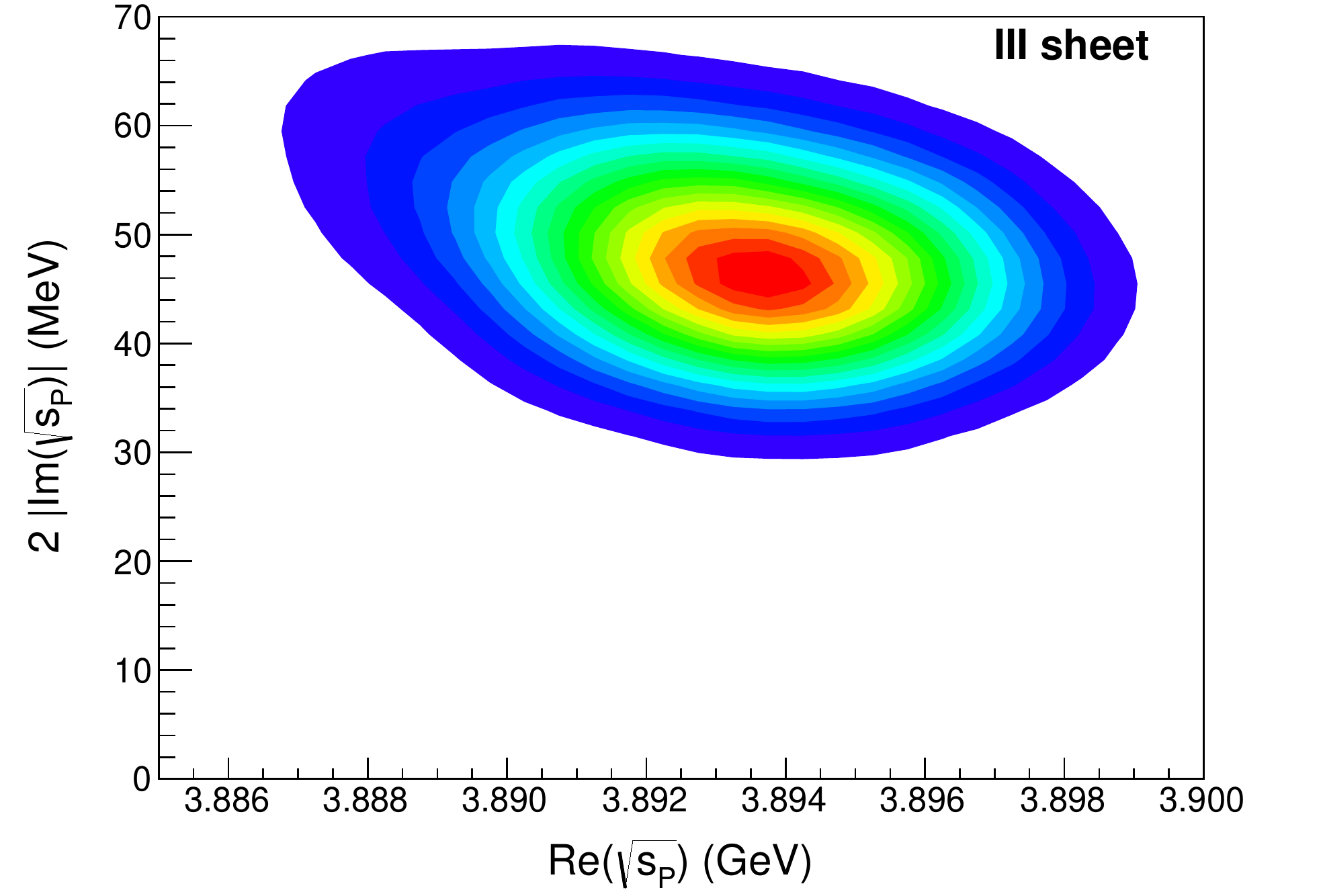}
\caption{Scenario \scenIII}
\label{fig:poleIII}
\end{subfigure}
 \begin{subfigure}[b]{.32\textwidth}
\includegraphics[width=\textwidth]{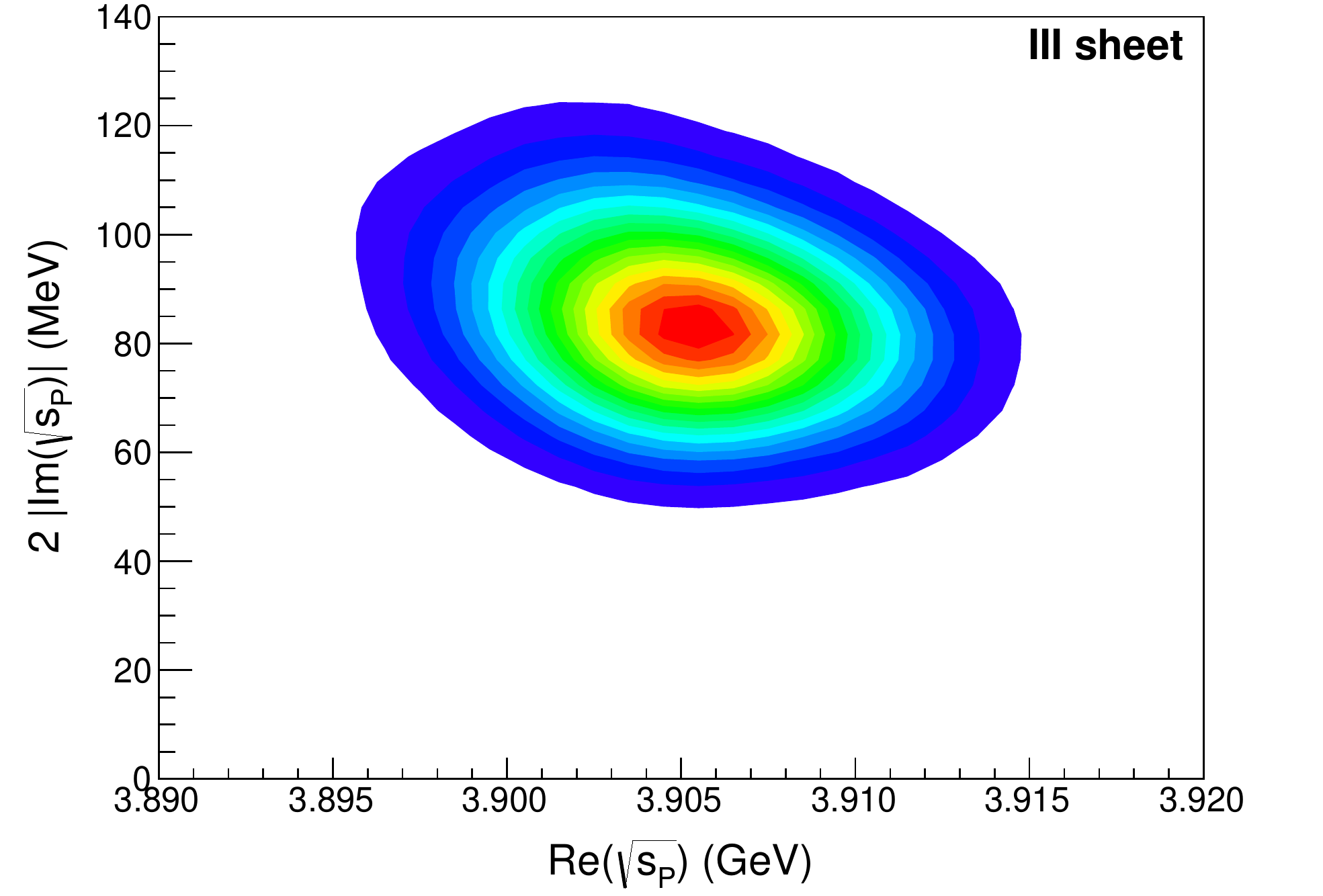}
\caption{Scenario \scenIIItr}
\label{fig:poleIIItr}
\end{subfigure}
 \begin{subfigure}[b]{.32\textwidth}
 \includegraphics[width=\textwidth]{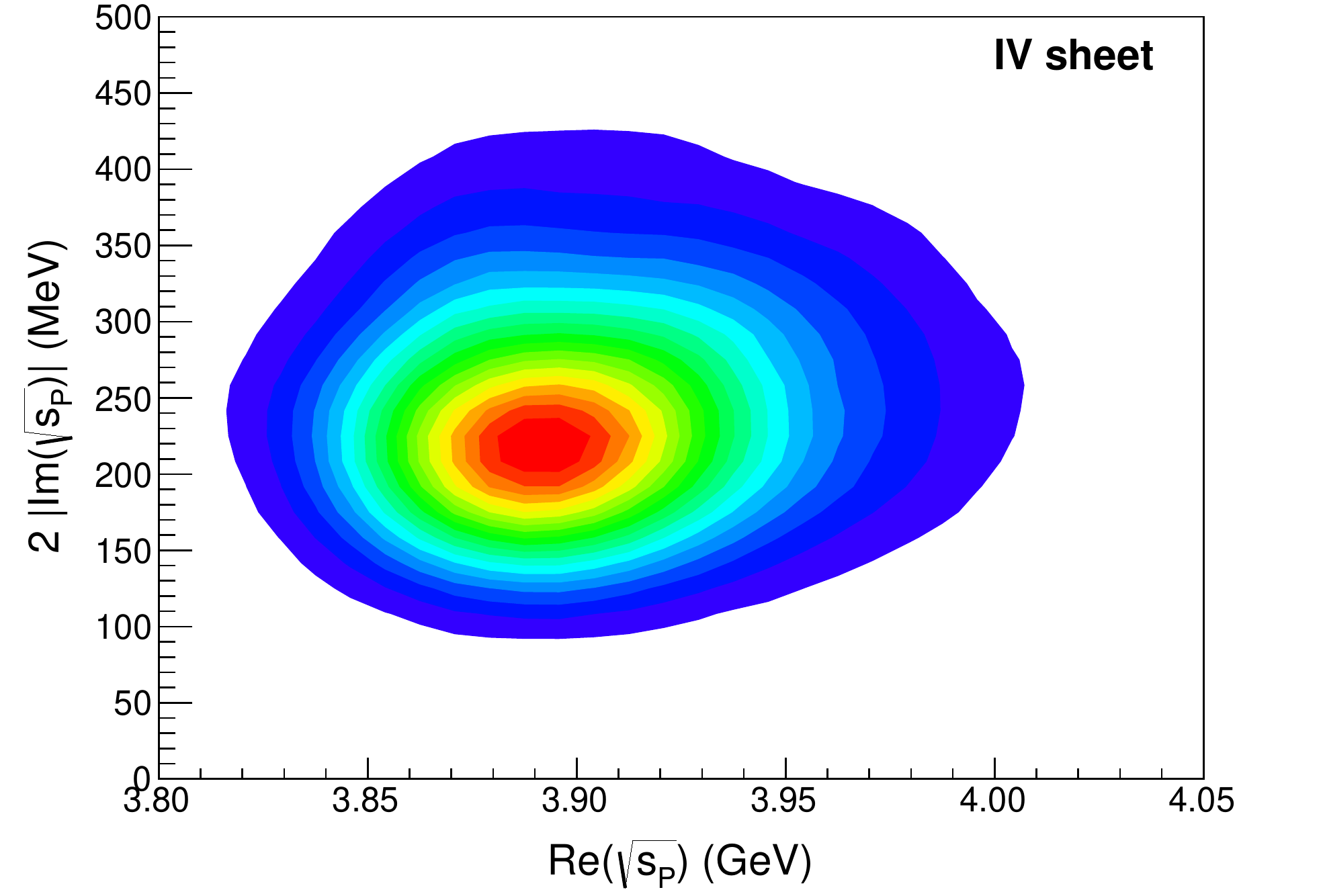} 
 \caption{Scenario \scenIV}
\label{fig:poleIV}
\end{subfigure}
 \caption{Pole position according to the scenarios which allow for the presence of a pole in the scattering matrix close to the physical region. The colored regions correspond to the $1\sigma$ confidence level. }
 \label{fig:poles}
 \end{figure}
The existence of a $Z_c$ state is equivalent to the appearance of a pole in the unphysical sheets of the scattering amplitude. As discussed in \sectionname{\ref{sec:amplitude}}, the Riemann sheet where the pole appears can give hints on its microscopic origin. For each one of the three scenarios that allow for the presence of a pole, we can calculate the pole position, and estimate its statistical uncertainty according to the bootstrap analysis we discussed in previous section. In \figurename{~\ref{fig:poles}} we show the pole position according to the $68\%$ fraction of best $\chi^2$ obtained in the bootstrap analysis. This can be translated into the $1\sigma$ region where the pole is expected to occur. The results are summarized in \tablename{\ref{tab:poles}}, and the main observations are as follows:
 \begin{enumerate}
 \item \scenIII: The pole appears above the $\bar D D^*$ threshold, on the III sheet (the closest to the physical region), and the width is $\Gamma \simeq 50\mev$. This is marginally compatible with the value quoted in the PDG, $M = 3886.6 \pm 2.4\mev$, $\Gamma = 28.1 \pm 2.6 \mev$~\cite{pdg}. The reasons for this slight discrepancy are twofold: $i)$ in the fits
  performed in the experimental analysis the sum of the signal (Breit-Wigner) and background (phase-space shaped) is performed incoherently, which tends to provide narrower values for the width; $ii)$ in particular for the $\jpsi\pi^0\pi^0$ data, we cannot disentangle the Breit-Wigner width from the experimental resolution, effectively giving a slightly larger width to the resonance. 
\item \scenIIItr:  The presence of the logarithmic branching point close to the physical region allows for the pole to be slightly deeper in the complex plane, with a width $\Gamma \simeq 90\mev$. The mass is still safely above threshold. 
\item \scenIV: In this case the peak is generated by the combination of the logarithmic branching point with the virtual state pole on the IV sheet. Given the presence of the triangle singularity, the position of the pole is not well constrained. The width, defined in analogy to the Breit-Wigner case as $2 \Im \sqrt{s_P}$,  is broader than in the other scenarios, $\Gamma \simeq 250\mev$, but the mass is unchanged, albeit with errors of $\sim 100\mev$.
\item \scentr:  By construction, this scenario does not allow for poles close to the physical region.
  \end{enumerate}
 
 \begin{table}[h]
\centering
\begin{tabular}{l|ccc}
 &  \scenIII & \scenIIItr & \scenIV \\ \hline
$M \equiv \Re \sqrt{s_P}$ (MeV) &  $3893.2^{+5.5}_{-7.7}$ & $3905^{+11}_{-9}$ & $3900^{+140}_{-90}$\\
$\Gamma \equiv 2 \left|\Im \sqrt{s_P}\right|$ (MeV)&  $48^{+19}_{-14}$  &$85^{+45}_{-26}$  & $240^{+230}_{-130}$ \\
\end{tabular}
\caption{Mass and width of the $Z_c(3900)$ according to the scenarios which allow for the presence of a pole. The error quoted is the $1\sigma$ statistical uncertainty obtained with the bootstrap analysis of \sectionname{\ref{sec:fits}}. } 
\label{tab:poles}
\end{table}
 
 \section{Conclusions}
The literature on $XYZ$ states abounds with discussions about their microscopic nature. In this letter we show how a thorough amplitude analysis can help in constraining the various different phenomenological models. We tested four different scenarios, corresponding to pure QCD states, virtual states, or purely kinematical enhancements. The best fit is obtained for a compact QCD state, but the rejection of the other scenarios is not significant. We conclude that given the present data, specifically mass projections, it is not possible to distinguish between the different hypotheses. Future high-statistics measurements and the study of the full Dalitz plot, thus including angular correlations, will improve the  discrimination power of our analysis, in particular by constraining the contribution of the $D_1$ exchange. This new information, together with a combined analysis of other reactions, \eg $Y(4260)\to h_c \,\pi\pi$ or photoproduction off protons,   will allow us to shed more  light on the nature of the  exotic charmonium sector.
 
All material will be gathered onto an interactive website which will available online~\cite{Mathieu:2016mcy,JPACweb}.

%%%%%%%%%%%%%%%%%%%%%%%%%%%%%%%%
%	Ack
%%%%%%%%%%%%%%%%%%%%%%%%%%%%%%%%
\section*{Acknowledgments}
AP thanks A.~Esposito for fruitful discussions and comments on the manuscript, and AS thanks R.~Mitchell for discussions of the \bes results. We also thank C.~Hanhart for useful exchanges at and after the PWA/ATHOS workshop. 
This material is based upon work supported in part by the U.S.~Department of Energy, Office of Science, 
Office of Nuclear Physics under contract DE-AC05-06OR23177. 
This work was also supported in part by the U.S.~Department of Energy under Grant DE-FG0287ER40365, 
National Science Foundation under Grants PHY-1415459 and PHY-1205019, and IU Collaborative Research Grant. CF-R work is supported in part by research grant IA101717 from PAPIIT-DGAPA (UNAM),
by CONACYT (Mexico) grant No. 251817, 
and by Red Tem\'atica CONACYT de F\'{\i}sica de Altas Energ\'{\i}as (Red FAE, Mexico).

\bibliographystyle{elsarticle-num}
\bibliography{quattro}

\end{document}

%% file: quattro-sym.tex
\usepackage{afterpage}
\usepackage{amsfonts}
\usepackage{amsmath}
\usepackage{amsthm}
\usepackage{amssymb}

\usepackage{bm}
\usepackage{bbm}
\usepackage{color}
\usepackage{dsfont}
\usepackage{graphicx}
\usepackage{hyperref}
\usepackage{mathbbol}
\usepackage{multirow}
\usepackage{relsize}
\usepackage{slashed}
\usepackage{subcaption}
\usepackage{url}
\usepackage{xspace}
\usepackage{bbding}

\let\Im\relax
\DeclareMathOperator{\Im}{Im} % this is how the logarithm should be denoted
\let\Re\relax
\DeclareMathOperator{\Re}{Re} % this is how the logarithm should be denoted
\newcommand{\XYZ}{\ensuremath{XYZ}\space}

\newcommand{\babar}{\mbox{\slshape B\kern-0.1em{\smaller A}\kern-0.1em
    B\kern-0.1em{\smaller A\kern-0.2em R}}\xspace}
\newcommand{\belle}{Belle\xspace}

\newcommand{\bes}{BESIII\xspace}

\newcommand{\sectionname}[1]{Section #1\xspace}

\renewcommand{\tablename}[1]{Table #1\xspace}

\newcommand{\evnospace}{\ensuremath{\mathrm{e\kern -0.1em V}}\xspace}
\newcommand{\mevnospace}{\ensuremath{{\mathrm{Me\kern -0.1em V}}}\xspace}
\newcommand{\ev}{\ensuremath{\mathrm{\,e\kern -0.1em V}}\xspace}
\newcommand{\kev}{\ensuremath{{\mathrm{\,ke\kern -0.1em V}}}\xspace}
\newcommand{\mev}{\ensuremath{{\mathrm{\,Me\kern -0.1em V}}}\xspace}
\newcommand{\gev}{\ensuremath{{\mathrm{\,Ge\kern -0.1em V}}}\xspace}
\newcommand{\gevcc}{\ensuremath{{\mathrm{\,Ge\kern -0.1em V}}/c^2}\xspace}
\newcommand{\tev}{\ensuremath{{\mathrm{\,Te\kern -0.1em V}}}\xspace}

\newcommand{\jpsi}{\ensuremath{{J\mskip -3mu/\mskip -2mu\psi\mskip 2mu}}\xspace}

\newcommand{\ie}{{\it i.e.}\xspace}

\newcommand{\eg}{{\it e.g.}\xspace}

\newcommand{\spect}[4][]{\ensuremath{\ifthenelse{\equal{#1}{}} {} {#1\,} {}^{#2\!} {#3}_{#4}}}

\newcommand{\DDstar}{\ensuremath{\bar D D^*}\xspace}

\def\Dbar    {\kern 0.2em\bar{\kern -0.2em D}{}\xspace}

\def\Dz      {\ensuremath{D^0}\xspace}
\def\Dzb     {\ensuremath{\Dbar^0}\xspace}
\def\DzDzb   {\ensuremath{\Dz {\kern -0.16em \Dzb}}\xspace}
\def\Dp      {\ensuremath{D^+}\xspace}
\def\Dm      {\ensuremath{D^-}\xspace}

\def\DpDm    {\ensuremath{\Dp {\kern -0.16em \Dm}}\xspace}

\def\Bbar    {\kern 0.2em\bar{\kern -0.2em B}{}\xspace}

\def\Lbar     {\kern 0.2em\overline{\kern -0.2em\Lambda\kern 0.05em}\kern-0.05em{}\xspace}

\newcommand{\scenIII}{III\xspace}
\newcommand{\scenIIItr}{III+tr.\xspace}
\newcommand{\scenIV}{IV+tr.\xspace}
\newcommand{\scentr}{tr.\xspace}